\documentclass[10pt,leqno]{amsart}
\usepackage{graphicx}
\usepackage{indentfirst,csquotes}
\RequirePackage{amsthm,amsmath,amsfonts,amssymb}
\usepackage{xcolor,paralist,titlesec,fancyhdr,etoolbox}

\titlespacing*{\section}{0pt}{2.0ex}{2ex}

\makeatletter
\renewcommand*{\@seccntformat}[1]{\csname the#1\endcsname\hspace{0.5cm}}
\makeatother

 \usepackage[foot]{amsaddr}
 
\RequirePackage[colorlinks,citecolor=blue,urlcolor=blue]{hyperref}

\usepackage[sort&compress]{natbib}

\usepackage{amssymb}
\usepackage{amsthm}
\newcommand{\R}{\mathbb{R}}

\newcommand{\Indicator}[1]{\mathbb{I}({#1 })}

\usepackage{seqsplit}
\newcommand{\mubf}{\boldsymbol{\mu}}

\usepackage{xcolor}

\newcommand{\red}{ \bf \color{red} }
\usepackage{multirow}
\usepackage{setspace}
\usepackage[toc,page]{appendix}
\AtBeginEnvironment{tabular}{\singlespacing}
\usepackage{changes}

\makeatletter
\def\subsection{\@startsection{subsection}{2}%
  \z@{.5\linespacing\@plus.7\linespacing}{.3\linespacing}%
  {\normalfont\bfseries}}
\makeatother

\begin{document}

\title{A Bayesian Methodology for Estimation for Sparse Canonical Correlation}

\begin{center}
 \maketitle
 
 \normalsize
  Siddhesh Kulkarni\textsuperscript{1,3}, Subhadip Pal\textsuperscript{2}, and
  Jeremy T.\ Gaskins\textsuperscript{3}  \par \bigskip
  \textsuperscript{1} Bristol Myers Squibb, 556 Morris Avenue, Summit, NJ, 07901, USA \par
  \textsuperscript{2} United Arab Emirates University, Sheik Khalifa Bin Zayed Street, Al Ain, Abu Dhabi, P.O.Box 15551, UAE\par 
  \textsuperscript{3} 485 E. Gray Street, Louisville, KY, 40202, USA
  \bigskip
  \today
\end{center}

\begin{abstract}
It can be challenging to perform an integrative statistical analysis of multi-view high-dimensional data acquired from different experiments on each subject who participated in a joint study. Canonical Correlation Analysis (CCA) is a statistical procedure for identifying  relationships between such data  sets. In that context, Structured  Sparse  CCA (ScSCCA) is a rapidly emerging methodological area that aims for robust modeling of the interrelations between the different data modalities by assuming the corresponding CCA directional vectors to be sparse. Although it is a rapidly growing area of statistical methodology development, there is a need for developing related methodologies in the Bayesian paradigm.  
In this manuscript, we propose a novel  ScSCCA approach
where we employ a Bayesian infinite factor model and aim to achieve robust estimation by encouraging sparsity in two different levels of the modeling framework.  Firstly, we utilize a multiplicative Half-Cauchy process prior to encourage  sparsity at the level of the latent variable loading matrices. 
Additionally, we promote further sparsity in the covariance matrix by using graphical horseshoe prior or diagonal structure.
We conduct multiple simulations to compare the performance of the proposed method with that of other frequently used  CCA procedures, and we apply the developed procedures to analyze multi-omics data arising from a breast cancer study. 
\end{abstract} 

\bigskip

\section{Introduction} 
Due to recent technological advances, it has become more common for researchers to acquire multi-view data pertaining to  multiple feature sets measured on each subject that participates in a joint study. A joint analysis 
considering the multiple views together can provide a more comprehensive understanding of the underlying phenomenon by facilitating the borrowing of information across the different views. Understanding the relationships between such multi-view data is challenging, especially in the case of  high-dimensional data-views that often occur in many modern areas of research including multi-omics studies. \par 
In order to study the interrelation between two data views,  Canonical Correlation Analysis \citep[CCA;][]{10.2307/2333955} has been a widely used statistical procedure.  It identifies the optimal projection directions for each view so that the corresponding projected  variables achieve the maximum Pearson correlation between them.  This key idea of CCA is to project the complicated high dimensional variables within each view to  low-dimensional latent spaces which are maximally correlated across views. CCA has been widely used to analyze such multi-view data sets, and many of its variants are successfully applied in various scientific areas of research in science including genomics \citep{witten2009extensions,waaijenborg2008quantifying}, computer vision \citep{lin2006frequency,zhang2013l1}, meteorology \citep{statheropoulos1998principal}, biomedicine \citep{li2009joint,zhang2014frequency}, imaging analysis \citep{lin2014correspondence, du2016structured}, among others. Interested readers are referred to \citet{yang2019survey} and \citet{zhuang2020technical} for a more
extensive review of this  area of research. 
\par 
Researchers often have  high dimensional data where the number of features $p$ measured on each subject is greater than number of subjects ($p \gg n$). This leads to inefficiency in traditional CCA due to overfitting. Sparse Canonical Correlation Analysis (SCCA) addresses this  by finding the meaningful features which contribute to the calculation of the canonical correlation. \par
In the frequentist approach to SCCA, there are two main strategies. One popular practice is to consider element-wise sparsity, which generally employs a penalty such as a lasso or fused lasso on the associated canonical loadings  \citep{parkhomenko2009sparse, witten2009extensions,waaijenborg2008quantifying, suo2017sparse}. 
%
  A second approach is Structured Sparse CCA (ScSCCA) where a given interconnection structure between the views is incorporated into the specification of the penalties on the canonical loadings  \citep{chen2012structured,lin2013group,lin2014correspondence}. For example, in the context of analyzing genetic data, the known biological relationships between genes can be used to construct such prior structural information.
    However,  one may not always have appropriate information to inform a prior structure about the relationships  across the different attributes, and incorporating the partial prior knowledge into the corresponding analysis is nontrivial. This has led to the development of ScSCCA approaches that uses graph/network guided fused lasso penalties \citep{chen2012structured, chen2013structure, yan2014transcriptome,du2015gn, chen2012efficient}.   
   
   
 An additional approach to CCA is based on  Interbattery Factor Analysis (IBFA). 
 The IBFA model \citep{tucker1958inter} decomposes the covariance for a particular view into a shared factor structure and the view-specific noises, and the dependence between views is determined by these shared latent factors. The covariance of the noise/residuals terms is typically considered as a diagonal matrix whose elements are sometimes referred to as the specific variances \citep{johnson2007applied}. In more realistic scenarios this assumption of independent noise among the features of same view may not hold and may prove inadequate to capture the interdependence among the features.  
 Many Bayesian approaches to CCA involve modeling based on the IBFA framework \citep{klami2007local,klami2013bayesian}, and we also follow this approach.
  \par
 To that end, we propose a new Bayesian method for the CCA problem, which performs joint estimation of canonical correlations as well as within-view covariance estimation by using sparsity-inducing priors. 
 A key contribution in this article is the  introduction of a novel infinite Bayesian factor shrinkage model through which we gain shrinkage in the loading matrices of the IBFA. 
Specifically, we utilize a multiplicative Half-Cauchy Process that facilitates flexible and adaptive dimension reduction for the factor loading coefficients. We also estimate the within-view covariances using the graphical horseshoe prior  \citep{li2019graphical} to encourage a flexible but sparse structure for these dependencies.
 We adopt a fully Bayesian approach that facilitates posterior inference and uncertainty quantification via Markov chain Monte
Carlo (MCMC) samples from the corresponding posterior distribution. 
 \par
 The rest of the article is organized as follows. 
 First,
 we review the mathematical formulation of  CCA. In Section \ref{section:model} we introduce the proposed Bayesian model and provide details on the choices of priors.  Section \ref{section:posterior_inference} includes the details of the MCMC algorithm that we have designed to explore the corresponding posterior distribution, and we discuss some of the challenges to performing posterior inference.  Section \ref{section:simulation} consists of comparisons across  competing estimation approaches in different simulation settings. In section \ref{section:datanalaysis} we analyze a set of breast cancer data to investigate the  relationships between the copy number and gene expression. Section \ref{section:conclusion} summarizes the project and discusses future directions.
 
\section{Factor Model Formation of CCA}\label{section:factormodel}
\subsection{Canonical Correlation Analysis}
In this section, we provide a few mathematical and technical details of CCA and discuss a commonly used factor model that can be used to conduct CCA. While we will focus on the two-view form of CCA,  the developed theory can be applied for a multi-view setup as we discuss in Section \ref{section:conclusion}.
We begin by specifying  notation.  We denote  scalar, vector, and matrix parameters by lowercase Roman, lowercase bold, and uppercase bold letters, respectively. Let $\mathbf{X}^{(1)} \in \R^{n \times p^{(1)}}$ and $\mathbf{X}^{(2)} \in \R^{n \times p^{(2)}}$ be the full data matrices corresponding to two data-views.
The sample size is given by $n$, and $p^{(m)}$ represents the dimensionality of each data-view ($m=1,2$). In this subsection  we assume that all the features are centered without loss of generality; that is, $E[\boldsymbol{x}_{i.}^{(m)}]=\mathbf{0}$ for $i=1, 2,\ldots, n$ and $m=1,2$.
The matrix  $\mathbf{\Sigma}$ represents the joint covariance over both data-views $\boldsymbol{x}_{i.}^{(1)}$ and $\boldsymbol{x}_{i.}^{(2)}$ and can be represented using the following block-matrix structure
\begin{equation}\label{Cov}
\mathbf{\Sigma}=
\begin{bmatrix}
\mathbf{\Sigma}^{(11)} & \mathbf{\Sigma}^{(12)}\\
\mathbf{\Sigma}^{(21)} & \mathbf{\Sigma}^{(22)}
\end{bmatrix}.
\end{equation}
Here, $\mathbf{\Sigma}^{(11)}$ and $\mathbf{\Sigma}^{(22)}$ represent the covariance matrices for the first and the second data-views, respectively, while $\mathbf{\Sigma}^{(12)}=\mathbf{\Sigma}^{(21)^T}$ denotes the covariance between the two data-views.
\par
 CCA aims to find the  vectors $\mathbf{u} \in \mathbb{R}^{p^{(1)}}$ and $\mathbf{v} \in \mathbb{R}^{p^{(2)}}$ so that the Pearson correlation between the linear combinations $\boldsymbol{x}_{i.}^{(1)T}\mathbf{u}$ and $\boldsymbol{x}_{i.}^{(2)T}\mathbf{v}$ is maximized. Hence, $\mathbf{u}$ and $\mathbf{v}$ are the feature loading values  for each of the two data-views that produce these linear combinations.
A formulation of the mathematical optimization corresponding to the standard CCA is  given as 
\begin{equation*}\label{CCA equation 1}
\rho={\mathrm{\text{max}} }\left\{\frac{\mathbf{u}^{T}\mathbf{\Sigma}^{(11)-1/2}\mathbf{\Sigma}^{(12)}\mathbf{\Sigma}^{(22)-1/2}\mathbf{v}}{\sqrt{\mathbf{u}^{T}\mathbf{u}}\,\sqrt{\mathbf{v}^{T}\mathbf{v}}} : \mathbf{u}  \in \mathbb{R}^{p^{(1)}}, \mathbf{v} \in \mathbb{R}^{p^{(2)}}\right\}
.\end{equation*}
This constrained optimization problem can be reformulated  as
\begin{equation}\label{optimization}
\rho= \underset{\mathbf{u}^*,\mathbf{v}^*}{\mathrm{\text{max}} }\left\{ \mathbf{u}^{*T}\mathbf{\Sigma}^{(11)-1/2}\mathbf{\Sigma}^{(12)}\mathbf{\Sigma}^{(22)-1/2}\mathbf{v^*} : \mathbf{u}^* \in \mathbb{S}^{p^{(1)}-1},\mathbf{v}^* \in \mathbb{S}^{p^{(2)}-1}\right\}. 
\end{equation}
Note that $\mathbb{S}^{p-1}=\left\{\boldsymbol{x} \in \mathbb{R}^{p} : \sqrt{\boldsymbol{x}^{T}\boldsymbol{x}} =1\right\}$ is the compact manifold of the set of $p$-dimensional vectors with norm 1. $\mathbf{u}^*$ and $\mathbf{v}^*$ are called the first canonical loadings and represent  the directions for each data-view in which the first canonical correlation is  maximized. Hence in this article, we will use the term canonical loadings and direction vectors interchangebly. 

The primary focus for inference is often  the triplet $(\rho, \mathbf{u}^*, \mathbf{v}^*)$, representing the first canonical correlation and its associated direction vectors. Additionally, there is sometimes interest in higher order canonical correlation terms which represent the next most impactful areas of dependence between the data-views. 
After finding the $(r-1)^{th}$ triplet $(\rho^{(r-1)}, \mathbf{u}^{*(r-1)}, \mathbf{v}^{*(r-1)})$, the $r^{th}$ set of canonical correlation parameters are obtained as
\begin{eqnarray}\label{optimization2}
&\rho_r= \underset{\mathbf{u}^*,\mathbf{v}^*}{\mathrm{\text{max}} }\left\{  \mathbf{u}^{*T}\mathbf{\Sigma}^{(11)-1/2}\mathbf{\Sigma}^{(12)}\mathbf{\Sigma}^{(22)-1/2}\mathbf{v^*} : \mathbf{u}^* \in \mathbb{S}^{p^{(1)}-1},\,\mathbf{v}^* \in \mathbb{S}^{p^{(2)}-1},  \right. \\
& \hspace{7em}  \left. \mathbf{u}^{*T} \mathbf{\Sigma}^{(11)}\mathbf{u}^{*(j)}=0,\,
\mathbf{v}^{*T} \mathbf{\Sigma}^{(22)}\mathbf{v}^{*(j)}=0, \, (j=1,\ldots,r-1)
\right\} . \nonumber
\end{eqnarray}
In order to perform the optimization corresponding to \eqref{optimization2}, the algorithm aims to find $\mathbf{u}^{*(r)}$ and $\mathbf{v}^{*(r)}$ that maximize the remaining correlation between the views while yielding linear combinations $\boldsymbol{x}_i^{(1)T}\mathbf{u}^{*(r)}$ and $\boldsymbol{x}_i^{(2)T}\mathbf{v}^{*(r)}$ that are uncorrelated with the previous $r-1$ linear combinations.
\par

With an estimate of the joint covariance matrix, one typically obtains the estimated canoncial correlations and direction vectors  through a singular value decomposition (SVD) of $\mathbf{\Sigma}$ 
\citep{yang2019survey}. Briefly, under SVD  the $p^{(1)} \times p^{(2)}$ matrix $\mathbf{\Sigma}^{(11)-1/2}\mathbf{\Sigma}^{(12)}\mathbf{\Sigma}^{(22)-1/2}$ 
from  \eqref{optimization} is decomposed as $\mathbf{LPQ}^{T}$. $\mathbf{P} \in \R^{p^{(1)} \times p^{(2)}}$ is a diagonal matrix of singular values, which after ordering give the canonical correlations $\rho_1,\rho_2,\ldots$.
The matrix $\mathbf{L}$ gives the left eigenvectors of $\mathbf{\Sigma}^{(11)-1/2}\mathbf{\Sigma}^{(12)}\mathbf{\Sigma}^{(22)-1/2}$, and hence, the $r$-th column provides the canonical loading vector $\mathbf{u}^{*(r)}\in\mathbb{S}^{p^{(1)}}$ for the first data-view.  Similarly, columns of $\mathbf{Q}$ provide the view 2 direction vectors $\mathbf{v}^{*(r)}$ for each canonical correlation.
As CCA is intimately related to the covariance between the two data-views, we will next introduce a latent factor model that facilitates estimation of this dependence. 

\subsection{Interbattery Factor Analysis}\label{latentfactormodel}
In this article we will use a version of IBFA proposed by \citet{bach2005probabilistics} for two data-view CCA.  This model uses loading matrices to project lower dimensional latent variables  to the higher dimensional space of the two data-views. 
The model is given as
\begin{eqnarray}\label{latentmodel}
\mathbf{z}_{i.} & \sim & \text{MVN}_{d}(0,\mathbf{I}),\nonumber \\ 
\boldsymbol{x}_{i.}^{(1)}\mid \mathbf{A}^{(1)}, {\mubf}^{(1)}, \mathbf{z}_{i.}, \mathbf{\Phi}^{(1)} & \sim & \text{MVN}_{p^{(1)}}({\mubf}^{(1)}+\mathbf{A}^{(1)}\mathbf{z}_{i.}, \mathbf{\Phi}^{(1)}),\nonumber \\
\boldsymbol{x}_{i.}^{(2)}\mid \mathbf{A}^{(2)}, {\mubf}^{(2)}, \mathbf{z}_{i.}, \mathbf{\Phi}^{(2)} & \sim & \text{MVN}_{p^{(2)}}({\mubf}^{(2)}+\mathbf{A}^{(2)}\mathbf{z}_{i.},\mathbf{\Phi}^{(2)}),
\end{eqnarray}
for $i =1, \ldots, n$, where $\text{MVN}_{d}({\mubf}, \mathbf{\Sigma})$ denotes a multivariate Normal distribution on $\R^{d}$ with mean $\mubf$ and variance-covariance matrix $\mathbf{\Sigma}$. 
Here, $d$ is the dimension of the latent variable $\mathbf{z}_{i.}$  which is less than the data dimensions ($\text{min} (p^{(1)}, p^{(2)}) \gg d$). In this model,  $\mathbf{A}^{(1)}$ and $\mathbf{A}^{(2)}$ project the lower dimensional latent space $\mathbf{z}_{i.}$ to the higher dimensional data spaces $\boldsymbol{x}_{i.}^{(1)}$ and $\boldsymbol{x}_{i.}^{(2)}$,
respectively. 
While we colloquially refer to these coefficient matrices as projection matrices, we note that they are not projections in the usual linear algebra sense.
The parameters $\mathbf{\Phi^{(1)}}$ and $\mathbf{\Phi}^{(2)}$ denote the within view covariance matrices representing variability beyond the factor structure.  Generally, these matrices are considered to be  diagonal, and the variances are referred to as the feature-specific variance.  For this study, we allow these matrices to have non-zero off-diagonal elements, so borrowing from the specific variance terminology, we will refer to the matrices $\mathbf{\Phi^{(1)}}$ and $\mathbf{\Phi}^{(2)}$ as a ``generalized specificity" for the view.
Integrating out the latent $\mathbf{z}_{i.}$ yields the marginal joint covariance matrix for $\boldsymbol{x}_{i.}^{(1)}$ and $\boldsymbol{x}_{i.}^{(2)}$
\begin{equation}\label{Grand_Cov}
\mathbf{\Sigma}=
\begin{bmatrix}
\mathbf{A}^{(1)}\mathbf{A}^{(1)T}+\mathbf{\Phi}^{(1)} & \mathbf{A}^{(1)T}\mathbf{A}^{(2)}\\
\mathbf{A}^{(2)T}\mathbf{A}^{(1)} & \mathbf{A}^{(2)} \mathbf{A}^{(2)T}+\mathbf{\Phi}^{(2)}
\end{bmatrix},
\end{equation}
with the 
same
block structure as in  \eqref{Cov}.
When $\mathbf{\Phi}^{(m)}$ are diagonal, the above matrix factorization substantially reduces the number of parameters to be estimated in 
$\mathbf{\Sigma}$
from $(p^{(1)}+ p^{(2)})^{2}/2$ to $(p^{(1)}+ p^{(2)})\times (d+1)$. Importantly, the covariance between the two data-views depends on the product of the two projection matrices,  $\mathbf{A}^{(1)T}\mathbf{A}^{(2)}$. Hence, the projection matrices are the critical parameters for the modeling of CCA. 
In the context of the latent factor model specified in $\eqref{latentmodel}$,
the CCA optimization problem in $\eqref{optimization}$ is written as 
\begin{equation}\label{optimization_modified}
\rho=\underset{\mathbf{u}^*\in \mathbb{S}^{p{(1)}} ,\mathbf{v}^* \in  \mathbb{S}^{p{(2)}}}{\mathrm{\text{ max}}}\left\{ \mathbf{u}^{*T}(\mathbf{A}^{(1)}\mathbf{A}^{(1)T}+\mathbf{\Phi}^{(1)})^{(-1/2)}\mathbf{A}^{(1)T}\mathbf{A}^{(2)}({\mathbf{A}^{(2)} \mathbf{A}^{(2)T}+\mathbf{\Phi}^{(2)}})^{-1/2}\mathbf{v^*} \right\}.\end{equation}
Consequently,  the optimization and the estimands ($\rho, \mathbf{u^*}, \mathbf{v^*}$) are a function of the projection matrices $\mathbf{A}^{(m)}$ and the generalized specificity matrices $\mathbf{\Phi}^{(m)}$. 
\par
It is important to note that as in a typical factor model, identifiablity issues arise for the parameters $\mathbf{A}^{(m)}$ if no  further constraints are considered. One can specify any semi-orthogonal matrix $\mathbf{O}$ such that $\mathbf{OO}^{T}=\mathbf{I}_{d \times d}$ and obtain $\tilde{\mathbf{A}}^{(m)}=\mathbf{A}^{(m)}\mathbf{O}$. Substituting $\tilde{\mathbf{A}}^{(m)}$  in \eqref{Grand_Cov}, the overall  $\mathbf{\Sigma}$ is unaffected, showing that the IBFA model parameters are non-identifiable. 
Although imposing  identifiablity constrains, such as forcing  $\mathbf{A}^{(m)}$ to be  lower triangular, can resolve the issue, this will also  induce an order dependence among the features. Alternatively,  specialized structures can be imposed to assign a special role to a few features. However, such restrictions  are not easy to generalize and often require domain specific expertise to choose an appropriate structure \citep{carvalho2008high, zhao2016bayesian}.
\par
However, it is important to note that the unidentifiablity of the projection matrices does not impact CCA  because the covariance matrix $\mathbf{\Sigma}$
is indentifiable, and the CCA estimands are functions of $\mathbf{\Sigma}$ \citep{bhattacharya2011sparse, geweke1996measuring}. Rather than imposing structural zeros in the lower-diagonal portion of the projection matrix, sparsity inducing priors can further support model stability and circumvent  issues related to non-identifiablity. \par

In addition to identifiability, another frequent issue to consider in the IBFA is  the number of latent factors $d$ to be used. A common practice is to fit the model with different choices of $d$ and then use a model selection criteria to obtain the optimal choice. Alternatively, in a Bayesian setting, it is also possible to consider a distribution on $d$ and obtain posterior samples through reversible jump MCMC \citep{lopes2004bayesian, miller2018mixture, yang2018fast}.  However, these algorithms tend to mix poorly in practice.  
An alternative approach is to  intentionally overfit the factor model, as introduced by  \citet{bhattacharya2011sparse} for Gaussian linear factor models. 
In their approach the number of factors $d$ is allowed to diverge to infinity, but 
 as additional factors are introduced to the model, they  play a progressively smaller role  in explaining the structure of the data.
 As shown in the next section, this is the approach we  take with our model.
 \par

 \section{Factor Shrinkage Model for Canonical Correlation Estimation}\label{section:model}
 \subsection{Non-Diagonal Factor Shrinkage Model (NDFSM)}
In this section we introduce our Non-Diagonal Factor Shrinking Model  for CCA, building on this IBFA framework to characterize the relationship between the data-views.  Our goal is to propose a model/prior structure  on the projection matrices $\mathbf{A}^{(m)}$ and the specificity matrices $\mathbf{\Phi}^{(m)}$ that  encourages stability in estimation by favoring sparsity and reducing the number of effective parameters.  This is paired with  modeling choices that also promote flexibility,  so that our methodology can adapt to the unique feature of the given data-views.

 \par 
Horseshoe priors \citep{carvalho2010horseshoe} belongs to a wide class of global-local shrinkage priors that are characterized by a local shrinkage parameter for recovering large signals and a global shrinkage parameter to adapt to the level of overall sparsity. 
In addition to their use in standard regression settings, they have also been used to encourage sparsity in latent factor models \citep[e.g.,][]{sekula2021single}. 
This class of global-local shrinkage priors exhibit a set of common features including heavy tails for robustness and appreciable mass near zero for inducing sparsity, leading to strong empirical performance.

In our model we use a horseshoe-like prior on each element of the loading matrices.
Our prior for $\mathbf{A}^{(m)} =\{ a_{jk}^{(m)}\}$  ($j=1,\ldots, p^{(m)}$;  $k=1,\ldots, d$;  $m=1,2$) has the following structure:
\begin{eqnarray}\label{projectionmatrixprior}
a_{jk}^{(m)} & \sim & \text{N} (0,\tau^{2(m)} \eta_k^{2} \lambda_{jk}^{2(m)}), \quad  \quad \nonumber \\
\lambda_{jk}^{(m)}  & \sim &  \text{C}^{+}(0,1), \nonumber \\
\tau^{(m)} & \sim & \text{C}^{+}(0,1), \nonumber \\
\quad \eta_k^{2} & = &  \prod_{j=1}^{k}\Tilde{\eta}_j^{2}, \nonumber \\ 
\Tilde{\eta}_j & \sim & \text{C}^{+}(0,\zeta),  \ (j>1); \ \ \Tilde{\eta}_1=1. \nonumber 
\end{eqnarray} 
Here $\text{C}^{+}(0,\zeta)$ represents a half-Cauchy random variable with density  $p(x) \propto (1+{x^2}/{\zeta^2})^{-1}\Indicator{x>0}$. The prior distribution for each  $a_{jk}^{(m)}$ of the loading matrix $\mathbf{A}^{(m)}$ has the variance $\tau^{2{(m)}} \eta_k^{2} \lambda_{jk}^{2{(m)}}$.  This choice is similar to a horseshoe prior except that we consider the product of three  terms instead of the usual two.
Here, $\eta_k^{2}$ is a factor-specific shrinkage parameter  which controls the sparsity of the each column of loading matrix to adaptively the determine the number of factors that contribute to the model.  This term  is constructed as a running product of half-Cauchy variables. $\tau^{2(m)}$ is the view-specific shrinkage  which 
acts as a global shrinkage parameter controlling the variability of  all $p^{(m)}d$ coefficients in the entire data-view. In addition, we also introduce the hyper-local shrinkage parameter $\lambda_{jk}^{(m)}$ which accounts for element-wise variability in the loading matrix. The hyper-local parameter $\lambda_{jk}^{(m)}$ and global parameters $\tau^{(m)}$ follow a standard half-Cauchy prior. 
\par


To see how  $\eta_k^{2}$ is able to control the effective model dimension, we note the similarity to the multiplicative gamma process of \citet{bhattacharya2011sparse} for (one-view) Gaussian linear factor models \citep[see also,][]{durante2017note, legramanti2020bayesian}. In their model, the variance for the $k^{th}$ column of the factor coefficient matrix contains the term $\eta_k^{2}=  \prod_{j=1}^{k}\Tilde{\eta}_j^{2}$, where the individual $\tilde{\eta}_j$s follow a gamma distribution.  They allow the number of factors $d$ to diverge to infinity and note the necessary conditions for the gamma hyperparameters to ensure that these $\eta_k^2$s are  stochastically decreasing.

 In a similar spirit to their multiplicative gamma process, 
 we refer to our model structure as a multiplicative half-Cauchy process (MHCP). 
 As shown in Section A of the Supplementary Materials, the MHCP places increasing  probability that the shrinkage variance $\eta_k$ will be in neighborhood of zero as $k$ increases.  
 Hence, for larger $k$,  columns of $\mathbf{A}^{(m)}$ will be shrunk approximately to zero, effectively removing the factor from the model.
 As shown in Lemma 3 of the Supplementary Materials, the $\text{C}^+$ scale parameter $\zeta$ controls the rate at which the column importance decreases since  the median of $\eta_k$ is given by $\zeta^{k-1}$ ($k\geq 2$).  Thus, one should only consider  $\zeta<1$, and we use $\zeta=0.5$ in all of our  experiments.
 The model is an infinite multiplicative process when $d\rightarrow\infty$, but in practice one chooses a relatively large $d$ and investigates the behavior of $\eta_d$ to ensure that it is approximately zero. 
 \par

 As noted previously, there has been past work considering Bayesian latent factor models for CCA analysis through the  IBFA framework.  In particular,
\citet{wang2007variational} uses an automatic relevance determination (ARD) prior to find structure and sparsity in the loading matrices along with (non-sparse) inverse
Wishart priors on 
$\mathbf{\Phi}^{(1)}$ and $\mathbf{\Phi}^{(2)}$.
 \citet{klami2007local} follow a similar strategy with non-diagonal structures for $\mathbf{\Phi}^{(m)}$.
Similarly, \citet{klami2013bayesian} use  IBFA  with an ARD prior to impose structure in the loading matrices.  Importantly, this structure encourages a column $k$ to be active in both projections (inducing correlation between views) or in only one view (to induce correlation within the view independent of the other view).  As this ARD prior allows both within and across covariance, the authors argue that diagonal $\mathbf{\Phi}^{(m)}$ are sufficient in their models. \citet{zhao2016bayesian} 
consider an IBFA model with a 
three level regularization in terms of global, factor specific and local shrinkage. They use a normal scale mixture model and three parameter beta distribution to provide sparsity. The authors also assume diagonal structures for $\mathbf{\Phi}^{(m)}$.
\par


Having specified our sparse prior process for the $\mathbf{A}^{(m)}$ matrices, we consider the prior structure for the generalized specificity matrices $\mathbf{\Phi}^{(m)}$.  
Recall from \eqref{Grand_Cov} that $\mathbf{\Sigma}^{(m)}=\mathbf{A}^{(m)}\mathbf{A}^{(m)T}+\mathbf{\Phi}^{(m)}$, and so $\mathbf{\Phi}^{(m)}$ represents the covariance between features of the same view that is unexplained by the factor structure. 
Unlike most  previous  methods, we assume that the structure in $\mathbf{\Phi}^{(m)}$ is arbitrary, and we do not impose a diagonal restriction which would assume that the latent factors explain the entire dependence between features of the same data-view.  However, we do believe that this matrix is likely to be highly sparse as the majority of the structure should be captured by the shared factors, so a non-sparse prior such as inverse Wishart would be ineffective in this case.
\par

To that end, we apply the graphical horseshoe prior \citep{li2019graphical} on the inverse of the generalized specificity matrices  $\mathbf{\Phi}^{(m)-1} = \mathbf{\Omega}^{(m)}$ for $m=1,2$:
\begin{eqnarray}\label{Graphical Horseshoe}
\omega_{ii}^{(m)} & \propto & 1 \quad \text{(Flat Prior)}\nonumber \quad  i=1, \ldots, p^{(m)}, \\
\omega_{ij}^{(m)} & \sim &  \text{N}(0,\alpha_{ij}^{2(m)}\beta^{2}),  \quad \text{for }  i<j, \nonumber\\
\alpha_{ij}^{(m)} & \sim & \text{C}^{+}(0,1), \quad \text{for }i<j, \nonumber \\
\beta & \sim & \text{C}^{+}(0,1).
\end{eqnarray}
The graphical horseshoe model applies horseshoe priors on the off-diagonal elements of the precision matrix and an improper flat prior on the diagonal elements, all under the constraint that  $\mathbf{\Omega}^{(m)}$ 
is positive definite.
This obviously enforces a symmetry constraint  $\omega_{ij}^{(m)} = \omega_{ji}^{(m)}$ ($i,j=1,\ldots,p^{(m)}; i\ne j$). Despite utilizing the improper flat prior on $\omega_{ii}^{(m)}$, the positive definiteness constraint 
ensures that the full prior is proper \citep{li2019graphical}. 
For the individual $\omega_{ij}^{(m)}$ terms, the local  parameters $\alpha_{ij}^{(m)}$ adjust the magnitude of shrinkage for each term, while $\beta$ adapts to the sparsity of the entire matrix $\mathbf{\Omega}^{(m)}$ as a global shrinkage parameter. \par
We assume the mean-zero Gaussian prior for the view-specific mean vectors $\mubf^{(1)}$ and $\mubf^{(2)}$ through
\begin{eqnarray}
\mubf^{(1)} & \sim & \text{MVN}_{p^{(1)}}(0,\sigma^2\mathbf{I}), \nonumber \\
\mubf^{(2)} & \sim & \text{MVN}_{p^{(2)}}(0,\sigma^2\mathbf{I}), 
\end{eqnarray}
where $\sigma^2$ is a fixed hyperparameter which we typically set to 100.

\subsection{Diagonal Factor Shrinkage Model (DFSM) }\label{subsection:Digonal_shrinkagemodel}
As many CCA factor models use a diagonal structure for the generalized specificity matrices, we also choose to construct the analogous version of our NDFSM that uses a diagonal structure, called the   
Diagonal Factor Shrinking Model (DFSM).  This choice continues to use our multiplicative half-Cauchy prior for $\mathbf{A}^{(m)}$ and will produce sparse coefficient matrices.  As discussed previously, all dependence both within and between views will be determined through these projection matrices when restricting $\mathbf{\Phi}^{(m)}$ to be diagonal.\par
In the DFSM version of the model, 
we constrain the generalized specificity matrices  to be diagonal, $\mathbf{\Phi}^{(m)}=\text{diag}(\phi_{11}^{(m)},\ldots, \phi_{p^{(m)}p^{(m)}}^{(m)})$.  The specific variance $\phi_{jj}^{(m)}$ for feature $j$ in data-view $m$ is given a conjugate inverse gamma prior $\phi_{jj}^{(m)} \sim \text{IG}(0.1,0.1)$.
The rest of the model structure is the same as NDFSM.
\par

 

 
\section{Posterior Sampling and Inference}\label{section:posterior_inference}
\subsection{MCMC Algorithm for NDFSM}
We use an MCMC Gibbs sampling algorithm to draw posterior samples from our model. All the distributions are obtained through conjugacy.
The sampling algorithm iterates between the following steps.
\begin{enumerate}
\item Mean Vectors: We sample the full $(p^{(1)}+p^{(2)})$-dimensional mean vector $\mubf^{g}=({\mubf}^{(1) T}, {\mubf}^{(2) T})^{T}$ by marginalizing out the factor scores $\mathbf{z}_i$. 
Let $\mathbf{X}^g$ be the $n \times (p^{(1)}+p^{(2)})$ matrix of observations obtained by stacking two data-view matrices and $\mathbf{\Sigma}$ be the grand covariance matrix  \eqref{Grand_Cov} based on the current values of $\mathbf{A}^{(m)}$ and $\mathbf{\Phi}^{(m)}$. Let $\bar{\mathbf{X}}^g$ be the $(p^{(1)}+p^{(2)})$-dimensional vector of column means.
Then,
\begin{eqnarray*}
\mubf^{g}\mid \mathbf{\Sigma},\mathbf{X}^g & \sim & \text{MVN}_{(p^{(1)}+p^{(2)})} (\mubf^{*}, \mathbf{E}^{-1}), \nonumber \\  
\mathbf{E} & = & n\mathbf{\Sigma}^{-1}+\sigma^{-2}\mathbf{I}, \quad \nonumber \\ 
\mubf^{*} & = &\mathbf{E}^{-1}\mathbf{\Sigma}^{-1}\bar{\mathbf{X}}^g.   
\end{eqnarray*}       
\item Latent variable $\mathbf{z}_{i.}$: Each $\mathbf{z}_{i.}$ ($i=1,2,\ldots, n$) is sampled from
\begin{eqnarray*}
\mathbf{z}_{i.}\mid \mathbf{A}^{(1)}, \mathbf{A}^{(2)}, \mathbf{\Phi}^{(1)}, \mathbf{\Phi}^{(2)} & \sim &  \text{MVN}_{d}(\mubf^{*},\mathbf{E}^{*-1}),\nonumber \\
\mathbf{E}^{*} & = & \mathbf{I}_{d\times d}+\mathbf{A}^{(1)T}\mathbf{\Phi}^{(1)-1}\mathbf{A}^{(1)}+\mathbf{A}^{(2)T}\mathbf{\Phi}^{(2)-1}\mathbf{A}^{(2)}, \nonumber\\  
\mubf^{*} & = & \mathbf{E}^{*-1}\{\mathbf{A}^{(1)T}\mathbf{\Phi}^{(1)-1}(\boldsymbol{x}_{i.}^{(1)}-{\mubf}^{(1)} )+ \nonumber\\
&& \quad \mathbf{A}^{(2)T}\mathbf{\Phi}^{(2)-1}(\boldsymbol{x}_{i.}^{(2)}-{\mubf}^{(2)})\}.
\end{eqnarray*}
\item Loading matrices $\mathbf{A}^{(1)}$ and $\mathbf{A}^{(2)}$:
To facilitate sampling from the posterior of projection matrices, we use the data augmentation structure of \citet{makalic2015simple} for sampling from a half-Cauchy distribution.
\begin{enumerate}
\item $\mathbf{A}^{(1)}$ and $\mathbf{A}^{(2)}$: For $m= 1,2$ and each row $j=1,\ldots,p^{(m)}$, we sample 
\begin{eqnarray*}
\mathbf{a}^{(m)}_{j.}\mid \mathbf{\Phi}^{(m)}, \mathbf{Z}, \mathbf{X}^{(m)}, \mubf^{(m)} & \sim & \text{MVN}(\mubf^{**},\mathbf{E}^{** -1}),\nonumber \\
\mubf^{**}& = &  \tilde{\phi}_j^{-1(m)}\mathbf{E}^{**-1}\mathbf{Z}^{T}\mathbf{\Tilde{X}}^{(m)}_{.j}, \nonumber \\ 
\mathbf{E}^{**} & = & \tilde{\phi}_j^{-1(m)} \mathbf{Z}^{T}\mathbf{Z}+\mathbf{\Delta}^{-1}, \nonumber\\
\Tilde{x}_{ij}^{(m)} &=&  x_{ij}^{(m)}-\mu_{j}^{(m)}-[\mathbf{\Phi}_{j,-j}^{(m)}][\mathbf{\Phi}_{-j -j}^{(m)}]^{-1} \times \nonumber \\
&& \quad (\mathbf{X}_{i,-j}^{(m)}-\mubf_{-j}^{(m)}-\mathbf{A}^{(m)}_{-j.}\mathbf{z}_{i.}^{T}),\\
\tilde{\phi}_j^{(m)}&=& \mathbf{\Phi}_{j,j}^{(m)}-\mathbf{\Phi}_{j, -j}^{(m)}[\mathbf{\Phi}_{-j -j}^{(m)}]^{-1}\times \mathbf{\Phi}_{-j, j}^{(m)}.
\end{eqnarray*}
Here, $\mathbf{\Delta}$ is the $d\times d$ diagonal matrix of the shrinkage parameters, $\Delta_{kk}=\tau^{2(m)}\eta^{2}_k\lambda^{2(m)}_{jk}$ for $k=1,\ldots,d$.
The element $\Tilde{x}_{ij}^{(m)}$ from $\mathbf{\Tilde{X}}_{.j}$ is the data residual for observation $i$ after removing the effect of everything except the $j^{th}$ response variable in the $m^{th}$ view. 
Here, ${\Tilde{\phi}}_j^{(m)}$ is the variance of $j^{th}$ feature
conditionally on the other features.  
In the above we use the common shorthand where $\pmb{\Phi}_{ab}$ represents the sub-blocks of the matrix $\pmb{\Phi}$ given by rows $a$ and columns $b$; $j$ indicates that only the $j^{th}$ row/column is included and $-j$ denotes that all rows/columns except for the $j^{th}$ are included.
\item Hyperlocal Shrinkage parameters $\lambda_{jk}^{2(m)}$: For $m=1,2$; $j=1, \ldots, p^{(m)}$; $k=1,  \ldots, d$, we draw
\begin{equation*}
\lambda_{jk}^{2(m)} \mid C_{jk}^{(m)},a_{jk}^{(m)},\eta_k \sim \text{IG}\Big(1,\frac{a_{jk}^{2(m)}}{2\tau^{2(m)}\eta_k^{2}}+\frac{1}{C_{jk}^{(m)}}\Big).
\end{equation*}
\item
View-Specific Shrinkage Parameter $\tau^{2(m)}$: For $m=1,2$, we sample
\begin{equation*}
\tau^{2(m)} \mid F^{(m)},\mathbf{A}^{(m)},\eta_k \sim 
  \text{IG} \biggl(\frac{(p^{(m)}\times d)+1}{2},\sum_{k=1}^{d} \sum_{j=1}^{p^{(m)}}\frac{a_{jk}^{2(m)}}{2\lambda_{jk}^2\eta_k^{2}} +\frac{1}{F^{(m)}}\Biggr).
  \end{equation*}
\item Column-wise Shrinkage Parameter $\eta_k^{2}$: 
For $j=2$ to $d$, 
\begin{eqnarray*}
       \Tilde{\eta_j}^2 &\mid& \mathbf{A}^{(m)},{\Tilde{\eta}}_{(-j)},\tau^{(m)} \sim \text{IG} \Big(\frac{(d-(j-1))[p^{(1)}+p^{(2)}]+1}{2}, \\
       && \sum_{m=1}^2 \sum_{k=j}^{d} \sum_{i=1}^{p^{(m)}}\frac{a_{ik}^{2(m)}}{2\lambda_{ik}^{2(m)}\tau^{2(m)}\prod_{\underset{k'\neq j}{k'=1}}^{k}\Tilde{\eta}_{k'}^{2}}  +
       \frac{1}{H_j}\Big).
\end{eqnarray*}
After updating all $\Tilde{\eta_j}$,  we compute $\eta_k^{2} =  \prod_{j=1}^{k}\Tilde{\eta_j}^{2}$.
\item Data augmentation parameters $F^{(m)}$, $C_{jk}^{(m)}$, $H_j$: For $m=1,2$; $j=1, 2, \ldots p^{(m)}$ and $k=1, 2, \ldots, d$: 
\begin{eqnarray*}
C_{jk}^{(m)}\mid \lambda_{jk}^{(m)} & \sim & \text{IG}\Big(1,1+\frac{1}{\lambda_{jk}^{2(m)}}\Big),
\nonumber \ \\
F^{(m)}\mid \tau^{(m)} & \sim & \text{IG}\Big(1,1+\frac{1}{\tau^{2(m)}}\Big) ,
\nonumber\\
H_j \mid \eta & \sim &  \text{IG}\Big(1,\frac{1}{\zeta^2}+\frac{1}{\Tilde{\eta_j}^{2}}\Big).
\end{eqnarray*}
\end{enumerate}
\item Generalized Specificity Matrices $\mathbf{\Phi}^{(1)}$ and $\mathbf{\Phi}^{(2)}$: Under the  NDFSM with the GHS prior, we simply follow the sampling schemes described in 
\citet{li2019graphical} based on the sample sum of square matrix  
\begin{eqnarray}
\mathbf{S}^{(m)}=\sum_{i=1}^n (\boldsymbol{x}_{i.}^{(m)}-{\mubf}^{(m)}-\mathbf{A}^{(m)}\mathbf{z}_{i.})^{T}(\boldsymbol{x}_{i.}^{(m)}-\mubf^{(m)}-\mathbf{A}^{(m)}\mathbf{z}_{i.}).\nonumber
\end{eqnarray}
\end{enumerate}

\subsection{MCMC Algorithm for DFSM}
For the  version of our model that uses the diagonal specificity matrix, MCMC sampling is largely the same as in the NDFSM model.  To obtain posterior samples from the DFSM model, we replace step 4 of the previous algorithm with a corresponding conjugate sampler. For each $j=1,\ldots,p^{(m)}$ and $m=1,2$, we draw  $\phi^{(m)}_{jj} \sim \text{IG} (0.5n+0.1, 0.1+0.5\sum_{i=1}^{n}\Tilde{{x}}_{ij}^2)$. As before, the relevant  residuals are $\tilde{\boldsymbol{x}}_{i.}^{(m)} = \boldsymbol{x}_{i.}^{(m)}-{\mubf}^{(m)}-\mathbf{A}^{(m)}\mathbf{z}_{i.}$.
\subsection{Point Estimation and Inference}\label{subsection:Point estimate and inference}
We use the previously described Gibbs samplers  to obtain a large number of posterior samples from our model.  The main parameters required for CCA inference are the loading matrices $\mathbf{A}^{(m)}$   and the within data-view covariances $\mathbf{\Phi}^{(m)}$.  These parameters determine the overall covariance structures among and across the data-views through \eqref{Grand_Cov}, and they determine the value of the canonical correlation $\rho$ and the direction vectors $\mathbf{u}^{*}$ and $\mathbf{v}^{*}$ through \eqref{optimization_modified}.  For each set of posterior samples of $\mathbf{A}^{(1)}, \mathbf{A}^{(2)}, \mathbf{\Phi}^{(1)}, \mathbf{\Phi}^{(2)}$, we can obtain a posterior sample of $\rho, \mathbf{u}^{*}, \mathbf{v}^{*}$, as well as any higher order canonical correlations.
\par
To evaluate the mixing of the  MCMC algorithm, we inspect the traceplots and autocorrelation plots of the CCs, the log-likelihood, and other model parameters. Large autocorrelation among the MCMC samples leads to higher uncertainty in estimation of parameters, and the effective sample size represents the number of independent samples that would contain an equivalent amount of information as the correlated samples from  the given MCMC output. In the current context,  we seek to run the MCMC long enough to obtain an effective sample size of at least 1000 for the key parameters of interest.\par
To obtain point estimates, we estimate the canonical correlations by taking the  sample mean over the  $\rho$s sampled from each MCMC iteration (after burn-in and thinning).
Similarly, at each iteration we calculate the direction vectors 
from the SVD of $\mathbf{\Sigma}^{(11)-1/2}\mathbf{\Sigma}^{(12)}\mathbf{\Sigma}^{(22)-1/2}$
(Section \ref{section:factormodel}). This type of decomposition is not unique as the vectors could be reflected across the origin. To ensure that the canonical loadings are ``pointing'' in the same direction across MCMC samples, we first take the mean absolute values across all iterations for each element of the canonical loadings for both data-views and select the feature (from either data-view) with the largest absolute loading. 
We  impose an identifiability constraint on this feature to ensure that it maintains a positive sign in all iterations, so that all direction vectors are pointing in the same direction based on this influential feature.  
That is, if the loading for the selected feature is negative in a given iteration, we swap the signs of both $\mathbf{u}^*$ and $\mathbf{v}^*$ in that iteration to ensure that the correlation of $\boldsymbol{x}_{i.}^{(1)T}\mathbf{u^*}$, and $\boldsymbol{x}_{i.}^{(2)T}\mathbf{v^*}$ remains the same; if the loading for the selected feature is positive, we make no adjustment. After ensuring comparability across all iterations through this identifiability constraint, direction vectors $\hat{\mathbf{u}}^{*}$ and $\hat{\mathbf{v}}^{*}$ are obtained by averaging across iterations and dividing by the norm to ensure that they are unit $1$. 
\par
An important step in CCA is determining which features significantly load onto the direction vectors; that is, which elements of $\mathbf{u}^*$ and $\mathbf{v}^*$ are significantly different from zero.  To that end, we utilize a credible interval approach to determine significance.  Based on the identifiability-adjusted posterior samples of $\mathbf{u}^*$ and $\mathbf{v}^*$, we obtain a credible interval for each element of each vector and investigate whether or not it contains zero.
Recall that this orthonormal direction vectors are complex functions of the parameters $\mathbf{A}^{(m)}$   and $\mathbf{\Phi}^{(m)}$ which come from heavy-tailed horseshoe models. Consequently the posteriors for the elements of the direction vectors also tend to have heavy tails.  It has been shown in a variety of contexts that a 95\% credible interval under a heavy-tailed prior produces intervals that are overly wide and under-powered for hypothesis testing \citep{van2017uncertainty,li2019graphical}.
Following the advice of \citet{li2019graphical} in the context of covariance selection in their GHS model, we use a 50\% credible interval to determine if a feature is significantly loaded onto the direction vector. 
As we will show in the next section, this choice yields good  performance in our empirical studies.
\par

A final key inference question is which model should be used, either the general NDFSM that allows correlations between features of the same data-view through both the latent factor structure and the generalized specificity or the more restrictive DFSM that assumes independence of the features conditionally on the latent factors.  
As will be shown in the next section, we find that in some cases (particularly those with $p\gg n$) NDFSM may overshrink the projection matrices $\mathbf{A}^{(m)}$ relative to the shrinkage imposed on the specificity matrices $\mathbf{\Phi}^{(m)}$.  A consequence of this behavior is that  $\mathbf{A}^{(1)}\mathbf{A}^{(2)T}$ will tend to mainly contain zeros, and the canonical correlation $\rho$ will be very low.  Fortunately, this is easy to diagnosis by investigating the estimate of $\rho$ and can easily be corrected by instead using the DFSM.  In our experience DFSM avoids overshrinking $\mathbf{A}^{(m)}$ by imposing maximal shrinkage in $\mathbf{\Phi}^{(m)}$ through zeros in all off-diagonal elements.  
\par 

To determine which model should be used in a given data set, we recommend the following strategy.
First, run the NDFSM model and check for suspected  overshrinking  by considering $P\left\{\rho_{1} < 0.2\right\}>0.5$.  That is, if  the event that the  first CC is less than 0.2 has probability greater than 0.5, then we believe overshrinkage may be happening, and instead the base inference of the DFSM output.  Note that this threshold of $0.2$ is somewhat ad hoc, and other users may prefer a different criteria for switching from the general NDFSM to the more constrained DFSM choice.

R code to run the NDFSM and DFSM samplers is available at \newline
\url{https://github.com/SiddheshKulkarni-source/CCA}.   
This includes code to run the Gibbs samplers as well as to evalute mixing and perform posterior inference.

\section{Simulations} \label{section:simulation}
\subsection{Simulation Settings} 
To validate our proposed methodology  and to compare it against some competing methods, we perform seven simulation experiments. For synthetic data generation we use the IBFA model as specified in  \eqref{latentmodel}. 
For each of the  seven  simulation settings, we analyze 100 replicated data sets,  based on  different settings of projection matrices and within view covariances as summarized in Table \ref{settings}.\par

\begin{table}[tb]
  \centering
   \caption{Simulation Settings}
   \label{settings}
 \begin{tabular}{ccccc}
    \hline
      & Sample  &   & Latent  &  \\
     Setting &  Size & Specificity Structure &  Factors & Canoncial Correlations \\
    \hline
    1 & $n=300$ & AR Dependence & $d=1$ & $\rho_1=0.73,\rho_2=0.00$ \\
    2 & $n=50$ & AR Dependence & $d=1$ &$\rho_1=0.73,\rho_2=0.00$ \\
     3 & $n=300$ & $\mathbf{\Phi}^{(m)}=\mathbf{I}_{p^{(m)}}$  & $d=1$ & $\rho_1=0.70,\rho_2=0.00$\\
     4 & $n=50$ & $\mathbf{\Phi}^{(m)}=\mathbf{I}_{p^{(m)}}$ & $d=1$ & $\rho_1=0.70,\rho_2=0.00$\\
     5 & $n=300$ & AR Dependence & $d=10$ & $ \rho_1=0.73, \rho_2=0.60$\\
     6 & $n=50$ & AR Dependence & $d=10$ & $\rho_1=0.73, \rho_2=0.60$\\
     7 & $n=300$ & AR Dependence & $d=1$ & $\rho_1=0.49, \rho_2=0.00$\\
     \hline
    \end{tabular}
\end{table}

The true $\mathbf{A}^{(m)}$ matrices are generated by setting elements $1,11,21$  in the first column of $\mathbf{A}^{(1)}$ to 1, while elements $1,11$ in the first column of $\mathbf{A}^{(2)}$ are set to 1 and -1, respectively; all other elements in this column are zero. This first column is responsible for determining the first CC value. The elements in the other columns of the projection matrices are non-zero with probability 0.05 and drawn from standard normal. 
In some settings, we use an autoregressive structure for a non-diagonal choice of the generalized specificity $\mathbf{\Phi}^{(1)}$ and $\mathbf{\Phi}^{(2)}$ (autocorrelations of 0.4 and 0.2, respectively), and in other settings we use an identity matrix.  In all cases, we use dimensions of $p^{(1)}=100$ and $p^{(2)}=50$ and consider $n=300$ ($n>p$)  and $n=50$  ($p>n$).  The mean vector is always zero.  We consider the true number of factors to be either $d=1$ for the case when  there is only one non-zero canonical correlation or $d=10$ so that there are ten non-zero canonical correlations.  
In the 7$^{th}$ setting, we introduce a ``scaling'' parameter that we  multiply all elements of projection matrix by. The scale is chosen to reduce the contribution of $\mathbf{\Sigma}^{(12)}=\mathbf{A}^{(1)T}\mathbf{A}^{(2)}$ in the decomposition  \eqref{Grand_Cov}, reducing the magnitude of the canonical correlation.
We provide the resulting first two canonical correlations $\rho_1$ and $\rho_2$ for each setting in Table \ref{settings}.

For every dataset, we  obtain estimates $\hat{\rho}_1$ and $\hat{\rho}_2$ of the first two canonical correlations, as well as estimates of the direction vectors $\hat{\mathbf{u}}^*$ and $\hat{\mathbf{v}}^*$ for the first canonical correlation through our proposed inference scheme.  
We measure the estimation accuracy of a canonical correlation as the root mean squared error (RMSE) between the true and estimated values:
\begin{equation*}
RMSE(\hat{\rho}_l,\rho_l)=\sqrt{\frac{1}{N}\sum_{i=1}^{N}(\hat{\rho}_{li}-\rho_l)^{2}\,} \ . 
\end{equation*} 
Here $\rho_l$ and $\hat{\rho}_{li}$ are the $l^{th}$ canonical correlation and its estimate in $i^{th}$ dataset ($l=1,2$), and $N=100$ is the total number of data sets analyzed. 
We also consider the average bias of the CC estimates by considering the average difference between estimate and the true value.
As the canonical loadings vectors are in the unit space, we measure accuracy as a root mean error based on one minus the cosine similarity between of two unit vectors.  We refer to this as root mean cosine error (RMCE) and compute it as
\begin{equation*}
RMCE(\hat{\mathbf{u}}^{*}, \mathbf{u}^{*})=\sqrt{\frac{1}{N}\sum_{i=1}^{N}(1-\hat{\mathbf{u}}^{*T}_{i}\mathbf{u})\,} \ .
\end{equation*}
$\mathbf{u}^{*}$ and $\hat{\mathbf{u}}^{*}_{i}$ are the  direction vector and its estimate in the $i^{th}$ dataset. 
We obtain $RMCE(\hat{\mathbf{v}}^{*},\mathbf{v}^{*})$ for the direction vector of the second view in an equivalent way.\par


\par

\subsection{Competing Methods}\label{subsection: Competing Methods}
For each of the 100 data sets generated according to the settings in Table \ref{settings}, we fit the data according to the following methods.  We compare our methodology to three frequentist and two Bayesian methods, as explained  below. \par
\begin{enumerate}
 \item NDFSM, DFSM: For both versions of our proposed models,  MCMC is run for 15,000 iterations with 5000 burn-in iterations in the low dimensional settings ($n=50$). Samples are thinned to store 2000 samples. In the higher dimensional settings ($n=300$), we run the model for 300,000 iterations with the first 50,000 discarded as burn-in.  A thinned sample of 5000 samples are stored.  Implementation is done in \verb!R!. On average the thinned samples  give an effective sample size of 1000--1200 for both the  first CC and the log-determinant of joint covariance matrix $\mathbf{\Sigma}$ from \eqref{Grand_Cov}.
 
\item NDFMS+DFSM: We include the combined strategy of selecting between NDFSM and DFSM for estimation.
As noted in Section \ref{subsection:Point estimate and inference}, we base inference on the NDFSM posterior samples unless these samples yield $P\left\{\rho_{1} < 0.2\right\}>0.5$.  In this case we suspect the NDFSM results may be impacted by overshrinkage and instead use the DFSM output for inference.
\item Bayesian Group Factor Analysis (GFA): The Bayesian Group Factor Model by \citet{klami2013bayesian} provides a Bayesian competitor to our approach that comes from a similarly constructed IBFA model. 
This method uses a combination of spike-and-slab and ARD priors for the projection matrices and a diagonal structure for  $\mathbf{\Phi}^{(m)}$.
Using  the ``GFA" package, we run MCMC for 600,000 iterations with 60,000 as burn-in and 2000 samples saved.  This gives an effective sample size of approximately 1000 for the first CC. We perform inference using these posterior samples using the same procedures described in Section \ref{subsection:Point estimate and inference}.\par
\item Graphical Horseshoe (GHS):  The GHS model \citep{li2019graphical} is directly applied to the $n \times (p^{(1)}+p^{(2)})$ joint data matrix $\mathbf{X}^g$ obtained by stacking both views.  This approach directly estimates the overall grand covariance $\mathbf{\Sigma}$ without considering any distinction between features of the different views.  Canonical correlations and direction vectors  are calculated from the posterior samples of the joint covariance matrix through \eqref{optimization} and other inference steps follow as with the other Bayesian methods.  We note that this choice of modeling the overall $\mathbf{\Sigma}$ covariance is not a common approach to performing CCA, but as the GHS imposes sparsity in $\mathbf{\Sigma}^{-1}$, it is conceivable that it can produce strong CCA estimates through its own form of regularization. We run MCMC  for 60,000 iterations with 5000 burn-in iterations. The sample is thinned to obtain 2000 samples yielding with effective sample size of approximately 1000.\par
\item Regularized CCA (RCCA): RCCA  \citep{vinod1976canonical} extends the regular CCA method for $p \gg n$ case by adding an $\ell_2$  penalty on the covariance matrix of each view. It is implemented by the \verb!R! package ``CCA'' \citep{gonzalez2008cca}.  
The regularization parameter is chosen using leave-one-out cross validation with the default choices in the \verb!estim.regul! function.
\item Sparse CCA (SCCA): We consider two implementations of SCCA  \citep{witten2009extensions}.  Firstly, a Lasso ($\ell_1$) penalty is applied to the canonical loadings. We refer to this model as SCCA (STD), denoting that this is the ``standard'' implementation of SCCA. The second method uses a  fused Lasso penalty  on the canonical loadings. This model is referred to as SCCA (O) with ``O" denoting ordered. Both methods are encoded in ``PMA" package. We use the \verb!CCA.optim! function to obtain optimal penalties for both the methods after which the CCA function is used to calculate the direction vectors and CCs.
\end{enumerate}

An additional Bayesian method that we do not consider in our simulation study is the  Bayesian group
factor Analysis with Structured Sparsity (BASS) model  by \citet{zhao2016bayesian}.  This model has similar goals 
and a similar modeling framework to the  GFA model  \citep{klami2013bayesian}. The authors  provide \verb!C++! codes (but not \verb!R!) for their methodology. Hence, we do not utilize that method for our simulation study, although based on the similarities between the model constructions, we would anticipate its performance to be similar to GFA.

\subsection{Simulation Results: Estimation of Canonical Correlations}
\begin{table}[tb]
  \centering
  \caption{Comparison Between RMSE of Different Methods for Estimation of 1st and 2nd CC}
    \begin{tabular}{cccccccc}
   \hline
      & \multicolumn{7}{c}{Simulation Setting} \\
Method & 1     & 2     & 3     & 4     & 5     & 6     & 7 \\
\hline          & \multicolumn{7}{c}{RMSE for 1st CC Estimation} \\
  \cline{2-8}
    NDFSM & \textbf{0.0220} & 0.2321 & \textbf{0.0218} & 0.0938 & \textbf{0.0218} & 0.1191 & 0.0535 \\
    DFSM  & 0.0299 & 0.0837 & 0.0220 & \textbf{0.0644} & 0.0304 & \textbf{0.0450} & 0.1451 \\
    NDFSM+DFSM & 0.0220 & 0.1118  & 0.0218 & 0.0644 & 0.0218 & 0.1087 & 0.0535 \\
    GFA   & 0.0371 & 0.2816 & 0.0479 & 0.1940 & 0.0362 & 0.2283 & 0.1867 \\
    GHS  & 0.0338 & \textbf{0.0573} & 0.0465 & 0.1282 & 0.0326 & 0.0534 & 0.0451 \\
    RCCA  & 0.1082 &	0.1647&	0.0868&	0.1794&	0.1057&	0.1674&	0.1203 \\
    SCCA (STD) &0.0507 & 0.1325 & 0.0553 & 0.1370 & 0.0567 & 0.1242 & 0.0833 \\
    SCCA (O) & 0.1154 & 0.0659 & 0.1462 & 0.1178 & 0.1148 & 0.1006 & \textbf{0.0323} \\
   \hline
          & \multicolumn{7}{c}{RMSE for 2nd CC Estimation} \\
  \cline{2-8}
    NDFSM & \textbf{0.0085} & \textbf{0.1112} & 0.0439 & 0.2313 & \textbf{0.0317} & 0.2883 & \textbf{0.0344} \\
    DFSM  & 0.1693 & 0.3663 & 0.0804 & 0.2112 & 0.0626 & 0.0624 & 0.1624 \\
    NDFSM+DFSM & 0.0085 & 0.1630 & 0.0439 & 0.2112 & 0.0317 & 0.2839 & 0.0344 \\
    GFA   & 0.2532 & 0.1165 & \textbf{0.0000} & \textbf{0.0000} & 0.0956 & 0.5083 & 0.2285 \\
    GHS   & 0.3610 & 0.5994 & 0.2431 & 0.4555 & 0.0349 & \textbf{0.0451} & 0.3561 \\
    RCCA  & 0.4291 &	0.8657&	0.4298 &	0.8476&	0.0842&	0.2792&	0.5162\\
    SCCA (STD) & 0.4293 & 0.7168 & 0.4433 & 0.6943 & 0.0474 & 0.1897 & 0.4173 \\
    SCCA (O) & 0.1200 & 0.7570 & \textbf{0.0000} & 0.7055 & 0.4368 & 0.1686 & 0.3971 \\
  \hline
    \end{tabular}%
  \label{Table:CCA Table }%
\end{table}%
We begin by comparing the estimation accuracy for the canonical correlation coefficients across the different methods in Table \ref{Table:CCA Table }.
Considering the first CC,  note that our NDFSM performs strongly across most cases.  It has the lowest error in cases 1,3 and 5, and its error is basically equivalent to the best model in case 7. In setting 4 DFSM performs the best, but it is also the true data generating model since $\mathbf{\Phi}^{(m)}$ is  the identity matrix.  In the high dimension cases with non-diagonal specificity (settings 2 and 6), NDFSM does not perform as well, although DFSM performs well in these cases. In case 2, GHS is the best performing model, and DFSM has similar RMSE.  The combined strategy substantially reduces the error rate of NDFSM in both of these settings; we  further investigate this in Section B.1 of the Supplementary Materials.  
GHS has consistently strong performance in estimating the first CC across all settings, and  GFA is consistently outperformed by our models.  Among the frequentist approaches,  the SCCA methods do fairly well although typically worse than the Bayesian models, and  RCCA tends to be slightly worse than SCCA.
\par
Turning to estimation of the 2nd CC in the lower half of Table \ref{Table:CCA Table }, we see that GFA has the strongest performance in settings 2--4 at correctly zeroing out the second CC, although it performs poorly in the other two $\rho_2=0$ cases (settings 1 and 7).  NDFSM has strong estimation of $\rho_2$, performing among the best in four  of the seven settings (not 3, 4 and 6); in these setting 4 and 6, most methods,apart from GFA, perform poorly and NDFSM is no worse than the majority.  In cases when NDFSM correctly captures $\rho_1$, it also correctly estimates $\rho_2$.  Conversely, GFA seems to perform better on $\rho_2$ than $\rho_1$, and despite the strong performance for the first canonical correlation, GHS has fairly large RMSE for the second CC (except in setting 6). Similarly, the two SCCA have reasonable estimation for $\rho_1$, but in cases 1--4 with $\rho_2=0$, they estimate a much larger second CC.
\par

\begin{table}[tb]
  \centering
  \caption{Comparison Between Average Bias of Different Methods for Estimation of 1st and 2nd CC}
    \begin{tabular}{cccccccc}
     \hline
    & \multicolumn{7}{c}{Simulation Settings} \\
     Method     & 1     & 2     & 3     & 4     & 5     & 6     & 7 \\
\hline          & \multicolumn{7}{c}{Average Bias for 1st CC} \\
    \cline{2-8}
    NDFSM & 0.0045 &  -0.1086 & 0.0028 & -0.0001 & 0.0042 & -0.0300 & -0.0043 \\
    DFSM  & -0.0191 & -0.0399 & -0.0001 & 0.0039 & -0.0201 & -0.0044 & -0.0853 \\
    NDFSM+DFSM & 0.0045 & -0.0485 & 0.0028 & -0.0001 & 0.0042 & -0.0262 & -0.0043 \\
    GFA   & 0.0619 & -0.0694 & 0.1245 & 0.0420 & 0.0568 & 0.3592 & 0.2164 \\
    GHS   & -0.0067 & -0.0228 & -0.0191 & -0.0704 & -0.0074 & 0.0062 & -0.0151 \\
    RCCA  & -0.1059  &0.1525&-0.0844&0.1694& -0.1019 &	0.1556&	0.0509 \\
    SCCA (STD) & -0.0169 & 0.0185 & 0.0024 & 0.0310 & -0.0243 & 0.0296 & 0.0267 \\
    SCCA (O) & -0.1030 & 0.0546 & -0.1338 & 0.0712 & -0.0996 & 0.0496 & -0.0040 \\
      \hline
          & \multicolumn{7}{c}{Average Bias for 2nd CC} \\
     \cline{2-8}
    NDFSM & 0.0069 & 0.1039 & 0.0429 & 0.2220 & -0.0029 & -0.2316 & 0.0308 \\
    DFSM  & 0.1670 & 0.3543 & 0.0781 & 0.1961 & -0.0498 & -0.0250 & 0.1575 \\
    NDFSM+DFSM & 0.0069 & 0.1414 & 0.0429 & 0.1972  & -0.0029 & -0.2276  & 0.0308 \\
    GFA   & -0.1318 & 0.0329 & 0.0000 & 0.0000 & 0.1538 & 0.2627 & 0.0201 \\
    GHS   & 0.3608 & 0.5987 & 0.2411 & 0.4506 & -0.0042 & 0.0303 & 0.3559 \\
    RCCA  &0.4288&	0.8623	&0.4296&	0.8445&	-0.0801&0.2692&	0.5035\\

    SCCA (STD) & 0.4182 & 0.6994 & 0.4327 & 0.6741 & 0.0020 & 0.1026 & 0.4070 \\
    SCCA (O) & 0.0525 & 0.7563 & 0.0000 & 0.6611 & -0.3523 & 0.1425 & 0.3949 \\
      \hline
    \end{tabular}%
  \label{Table:Bias Calculation}%
\end{table}%

In addition to the RMSE for $\rho_l$ estimation, we also consider the average bias in Table \ref{Table:Bias Calculation}. 
These results are generally consistent with the MSE results.  We see that our models are generally unbiased for $\rho_1$, although there is some evidence of bias for $\rho_2$ (positive when $\rho_2=0$ as in case 4 and negative if $\rho_2\ne 0$ as in case 6). As mentioned above, SCCA is clearly failing to penalize the higher order terms when $\rho_2=0$, yielding large estimates of this CC for all STD implementations and the ordered (O) cases when $p>n$.

In Section B.1 of the Supplementary Materials, we further investigate the combined inference strategy.  In particular, we note that the overshrinkage criteria is only activated in settings 2 and 6 and impacts relatively few data replications.  However, in these replications, the squared error loss is substantially reduced by using the DFSM output instead of the original NDFSM.

\subsection{Simulation Results: Estimation of First Direction Vector}

\begin{table}[tb]
  \centering
  \caption{Comparison Between RMCE of Different Methods For Estimation of Direction Vectors}
    \begin{tabular}{cccccccc}
   \hline
     & \multicolumn{7}{c}{Settings } \\
      Method    & 1     & 2     & 3     & 4     & 5     & 6     & 7 \\
\hline          & \multicolumn{7}{c}{RMCE for First Direction Vector of View 1} \\
  \cline{2-8}
    NDFSM & \textbf{0.0717} & 0.3780 & 0.0938 & 0.2780 & \textbf{0.0772} & 0.3491 & \textbf{0.1256} \\
    DFSM  & 0.1890 & 0.3466 & \textbf{0.0900} & \textbf{0.2717} & 0.1933 & 0.3397 & 0.5749 \\
    NDFSM+DFSM & 0.0717 & \textbf{0.3120} & 0.0938 & 0.2717 & 0.0772 & \textbf{0.3394} & 0.1256 \\
    GFA   & 0.4838 & 0.8116 & 0.4016 & 0.7204 & 0.5233 & 0.8650 & 0.8164 \\
    GHS   & 0.2742 & 0.4640 & 0.1625 & 0.4811 & 0.2869 & 0.4774 & 0.4970 \\
    RCCA  & 0.4386 &	0.8104&	0.9983&	0.7710&	0.4636&	0.8296&	0.9947 \\
    SCCA (STD) & 0.3134 & 0.7769 & 0.2966 & 0.8171 & 0.3501 & 0.8065 & 0.5116 \\
    SCCA (O) & 0.5342 & 0.8523 & 0.6602 & 0.7752 & 0.5526 & 0.8283 & 0.5166 \\
   \hline
          & \multicolumn{7}{c}{RMCE for First Direction Vector of View 2} \\
 \cline{2-8}
    NDFSM & \textbf{0.1004} & 0.3482 & 0.0936 & 0.2586 & \textbf{0.1029} & 0.3631 & \textbf{0.1497}\\
    DFSM  & 0.1096 & \textbf{0.2797} & \textbf{0.0885} & \textbf{0.2522} & 0.1216 & \textbf{0.2798} & 0.3292 \\
    NDFSM+DFSM & 0.1004 & 0.2978 & 0.0936 & 0.2586 & 0.1029 & 0.3567 & 0.1497 \\
    GFA   & 0.3596 & 0.7558 & 0.3247 & 0.6746 & 0.4153 & 0.7984 & 0.7261 \\
    GHS   & 0.1686 & 0.3858 & 0.1103 & 0.4294 & 0.1999 & 0.4076 & 0.3110 \\
    RCCA  & 0.3514 &	0.7586&	0.3601&	0.7142&	0.3846&	0.7825&	0.6857
 \\
    SCCA (STD) & 0.3732 & 0.7313 & 0.3877 & 0.7564 & 0.4350 & 0.7859 & 0.4544 \\
    SCCA (O) & 0.5678 & 0.8013 & 0.6290 & 0.7347 & 0.5501 & 0.7949 & 0.4213 \\
   \hline
    \end{tabular}%
  \label{Table:Direction Vectors}%
\end{table}%

In addition to the magnitude of the  canonical correlations, we frequently want  to consider the accuracy of estimation for the variable loadings associated with the first canonical correlation.
The root mean cosine errors for the estimated direction vectors in  view 1 and view 2 associated with the first canonical correlation are summarized in  Table \ref{Table:Direction Vectors}.\par

 This table shows that all versions of our models---DFSM, NDFSM, and the combined approach---have better performance than the other models. Beyond the relatively minor differences in the CC estimations, these more substantial improvements in the direction estimation  make our approach  a better alternative when we need to infer  the important contributing factors to the CC. 
That is, even in cases when the methods miss-state the magnitude of the relationship between views, NDFSM and DFSM tend to correctly find the combination and weights of features that determine this relationship.
  GHS and SCCA, which are quite competitive for the estimation of $\rho_1$, lag behind in this criteria by showing higher values of direction vector RMCEs.
 \par 
 
 \subsection{Simulation Results: Significant Variable Loadings}\label{Section:Significant_Variable_Loading}

In conjunction with the previous exploration of the accuracy of the canonical loading vectors $\hat{\mathbf{u}}^*$ and $\hat{\mathbf{v}}^*$, we also want to interrogate whether we are able to correctly detect whether a variable is significantly loaded or not.  
Recall that for Bayesian variable selection we consider a variable to be significantly associated with the CC direction if the 50\% credible interval for its factor loading excludes zero. Full results for this investigation can be found in the Supplementary Materials Section B.2, and we briefly describe a few key conclusions here.

The proposed NDFSM and DFSM methods are both effective at detecting the features that have substantial loading on the first principal component vectors, with each obtaining power greater than 88\% in six of the settings.  Compared to DFSM, NDFSM is better able  to  detect features that impact the CC through non-zero correlations in the generalized specificity, an ability that is expected from its non-diagonal model structure.  Both have low false discovery rates.  In our experiments, GFA has very high false discovery rates (40--60\% in most settings). In contrast, GHS has a low false discovery rate, but substantially reduced power relative to our  proposals.  Similarly, the penalized regression methods are unlikely to estimate a non-zero loading for a non-relevant feature, but they also tend to have lower power than our approach.  
In conclusion, our proposed models, NDFSM, DFSM, and the combined strategy, perform similarly or slightly better than most of the competing models for the estimation of first canonical correlation.  However, our models truly show a substantial benefit when considering the  estimation of the  direction vectors and the selection of the features that are significantly loaded. 
NDFSM consistently performs well if $n>p$, and  its occasional overshrinkage when $p>n$ can be mitigated by the combined strategy that switches to inference under the diagonal $\mathbf{\Phi}^{(m)}$.
In our experiments DFSM beats GFA, even though both are IBFA models with diagonal $\mathbf{\Phi}^{(m)}$.
GHS, while not designed for CCA, performs well in estimation of CCs but under-performs in estimation and variable selection of the  direction vectors. 
\section{Breast Cancer Data Application}\label{section:datanalaysis}

Joint analysis of multiple types of ``-omics'' data is an important part of modern biomedical research \citep{morris2017statistical,manzoni2018genome,castleberry2019integrated}. 
In particular, many studies  seek to understand the relationship between  gene expression and copy number variation \citep{hyman2002impact,pollack2002microarray}, and CCA has proved to be an effective tool for such analysis.
\par 

Breast Cancer is one of the most widely diagnosed types of cancer. It is the fifth 
most common
cause of cancer-related deaths with an estimated  2.26 million new cases worldwide
\citep{sung2021global} and is a leading cause of cancer deaths among women  \citep{ferlay2020global}.
The incidence rate of the breast cancer 
is affected by several epidemiological risk factors such as 
demographic characteristics, reproductive history, family history, and lifestyle factors \citep{momenimovahed2019epidemiological}. 
Additionally, there is ongoing research on the genetic factors associated with the risk of breast cancer.

\par
We apply our method to the breast cancer data described in \citet{chin2006genomic}. 
There are $n=89$ samples/observations on which both DNA and RNA data are available. 
As the first data-view, we consider the matrix of DNA copy numbers (DNA) for genes located on the $1^{st}$ chromosome, yielding $p^{(1)}=136$ responses per sample. 
The data also contains genetic expression levels (RNA) for 19,672 genes, and we construct the second data-view by including the 50 genes located on  chromosome 1 with the greatest variability in expression (based on interquartile range). We additionally  include  200  genes across the other 22 chromosome sites with the highest interquartile range, yielding $p^{(2)}=250$ in total. 
 Among these 250 genes, the 50 genes located on the first chromosome are anticipated to have a significant association with the first data-view, since they are located on the same chromosome. 
As SCCA (O) utilizes a fused lasso that depends on the feature ordering, we order the copy numbers in view 1 according to their chromosomal location,  and the RNA expressions in the second data-view are ordered according to their nucleotide position (within chromosome).
We standardize both data-views.
\par
Using these data-views, we apply the same set of methods utilized in Section \ref{subsection: Competing Methods}.  We run MCMC for 100,000 iterations after 25,000  burn-in iterations, and the sample is thinned to store 5000 samples. When  running the proposed NDFSM, we find evidence of the previously discussed  overshrinkage scenario.  Based on the results from the NDFSM MCMC output, we have a posterior probability that $\rho_1$ is less than 0.2 of 0.7349.  Hence, we consider inference based on the MCMC output from the DFSM model.  We note that these results do not indicate overshrinkage in DFSM since $\rho_1$ less than 0.2 has  approximately zero probability under DFSM. The 5000 stored  samples provide  an effective sample size of $2078$. For GFA, we run the MCMC for 600,000 iterations and retain a thinned sample of size 10,000 to obtain an effective sample size of $1797$. In the case of GHS, we ran the model for 20,000 iterations, saved 2000 samples which gave effective sample size of $2000$.
\par

The estimated first and second canonical correlations are shown in Table \ref{Table:Data_Analysis_Variable_Selection}, and there are substantial differences in the results  across the different methods. GFA  estimates  these two correlations to be  almost unity.  In fact, it estimates the ten largest canonical correlations to be greater than 0.95, indicating a much higher level of dependence between the data-views than is estimated in the other methods.
In contrast,  the SCCA approaches have substantially lower estimates between 0.60 and 0.65.  GHS and our proposed method yield similar estimates with $\hat{\rho}_1\approx 0.92$ and $\hat{\rho}_2\approx 0.90$.
\par


\begin{table}[tbp]
  \centering
  \caption{Canonical Correlation and Variable Selection Analysis for Breast Cancer Data}
    \begin{tabular}{p{15em}ccccc}
     \hline
     Method & NDFSM & GHS & GFA &  SCCA & SCCA \\
     & +DFSM &&& (O)& (STD)\\
     \hline
    Estimate of First CC & 0.9198 & 0.9387 & 0.9697 &  0.6071 & 0.6544 \\
    Estimate of Second CC & 0.9001 & 0.9106 & 0.9661 &  0.5599 & 0.6328 \\
    Sig.\ Copy Number Loading & 60    &{130}   & 1     &  59    & 18 \\
    Sig.\ Gene Loadings & {18}    & {0}     & 0     &  101   & 46 \\
     Sig.\ Gene Loadings & {8}     & {0}     & 0    &  26    & 11 \\
     \hspace{1em}on chromosome 1 &&&&& \\
    Weight (\%) of view 2 direction & {55.05} & {21.32} & 26.82 &  37.54 & 76.82 \\
    \hspace{1em}on chromosome 1 &&&&&\\
   \hline
    \end{tabular}%
  \label{Table:Data_Analysis_Variable_Selection}%
\end{table}%

 We also investigate the behavior of the estimated direction vectors and the number of significant loadings found in this analysis in Table \ref{Table:Data_Analysis_Variable_Selection}. 
 As compared to other Bayesian methods, the NDFMS+DFSM combination identifies 18  significant genes (view 2 components), while GHS and GFA do not identify any significant genes.
As we expect the genes located on chromosome 1 to be the most active features in determining the correlation to the chromosome 1 copy numbers, we investigate the sum of the square weights (both significant and non-significant) in the view 2 direction vector for the chromosome 1 genes. 
Since $1=\mathbf{v}^{*T}\mathbf{v^*}=\sum_{j=1}^{p^{(2)}}(v^{*}_j)^2$, the corresponding summation over only the 50 genes that are located on chromosome 1 can be viewed as a weight percentage for the contribution of the chromosome 1 genes to the canonical correlation.
 In Table \ref{Table:Data_Analysis_Variable_Selection} we can see that among the Bayesian methods, GHS and GFA assign 21\% and 27\% weight to the genes of the first chromosome, respectively. 
 In contrast, our model assigns 55\% of the loading weight to the genes of chromosome 1.
 \par

The frequentist method SCCA (O), which utilizes the chromosomal locations and nucleotide positions in the data, selects 101 significant genes for the first CC direction and 26 of these genes are located on chromosome 1.   That is, 26\% of significant genes are from chromosome 1 compared to 44\% (8 of 18) from NDFSM+DFSM.  The direction vector under SCCA (O) also assigns a lower weight to the chromosome 1 genes than does NDFSM+DFSM (38\% vs 55\%). 
Using the standard implementation of SCCA without the position information, 18 copy numbers are selected (fewer than NDFSM+DFSM and SCCA (O)) and 46 genes are selected (between the two methods) as significantly associated with the first CC.  While only 11 of these 46 genes are located on chromosome 1, these genes have very large loadings and account for 77\% of the direction vector.
\par

As the set of true population parameters are unknown for this real data, we cannot conclusively say which method provides the most accurate results. However, NDFSM+DFSM obtains estimates of the canoncial correlations that 
might be said to 
match the expected magnitude for this context.  More importantly, it estimates the largest portion of the significant loadings in view 2 to correspond to genes that are located on the chromosome corresponding the copy numbers in view 1; further, over half of the weight associated with the view 2 loadings come from the genes on the first chromosome.  These results support our claim that our methodology provides contextually reasonable conclusions for this data.


 \section{Conclusions}\label{section:conclusion} 
The rise of interconnected, high-dimensional, sparse data, measured on a small number of samples, demands the development of new statistical theory, and the ability to construct flexible and sparse models makes the Bayesian approach a valuable 
strategy for proposing new structured sparse CCA methodology.
Our model is one of  few which provide Bayesian modeling of within view covariance matrices with sparse CCA. To the best of our knowledge, ours is the only model which  models a sparse generalized specificity $\mathbf{\Phi}^{(m)}$ without restricting it to be diagonal. 
\par
As shown in the simulations, 
using GHS priors for $\mathbf{\Phi}^{(m)}$
 often performs better than GFA and DFSM which assume diagonal structures.  However, improved performance is not universal, and we observe cases where 
 NDFSM overshrinks the cross-covariance terms.  
The NDFSM+DFSM approach, which combines the GHS and diagonal versions of our models with an ad hoc selection rule, proves to be a competitive approach to diagnosis and correct this issue.
As overshrinkage  tends to only appear when $p>n$, one might also make the initial decision to only consider the DFSM choice when the sample size is low.
\par

While   CCA generally  considers  the case where 
there are 
only two views of the data, it can easily be generalized to a multi-view setting along the lines of 
\citet{zhao2016bayesian}. 
The IBFA model \eqref{latentmodel} is  extended by including 
regressions for each additional data-view as a function of the shared latent variables and 
the parameters $\mubf^{(m)}$, $\mathbf{A}^{(m)}$, and $\mathbf{\Phi}^{(m)}$.  Canonical correlations and direction vectors can be estimated for each pair $(m,m')$ of views.
Our proposed methodology can therefore be applied to this case by specifying our multiple half-Cauchy process as the prior for the additional $\mathbf{A}^{(m)}$ and a GHS prior for $\mathbf{\Phi}^{(m)}$.
\par

As part of our model specification, we have introduced a new flexible shrinkage prior, the multiplicative half-Cauchy process.  Along the lines of the multiplicative shrinkage processes \citep{bhattacharya2011sparse, schiavon2021generalized}, this model flexibly imposes shrinkage on the projection matrix coefficients, while  reducing the role of each subsequent factor.  
Beyond  the  results considered in the Supplementary Material Section A, it would  be of interest to further investigate the mathematical and theoretical features of this prior process.
However, this would be more straight-forwardly  achieved by considering its role within the context of a single-view factor model instead of our NDFSM.  The additional layer of multiple views and the non-diagonal residual variance complicates derivations, relative to exploration in a standard factor model.  We have run additional simulations (not shown) comparing our multiplicative half-Cauchy process to the multiplicative gamma process of \citet{bhattacharya2011sparse} and have found our approach to perform comparably and in some cases better.
\par

One of the main challenges of this model, and most Bayesian approaches to CCA, is computational scalability. 
As we use a completely Bayesian model and its corresponding Gibbs sampler, the estimation algorithm can be computationally intensive for higher dimensional problems. 
Approximate Bayes algorithms such as variational inference which seeks to find the posterior mode of an approximation distribution have been used previously with some success
\citep{klami2013bayesian, zhao2016bayesian}. 
While this may  find  parameter estimates more quickly than standard MCMC, these estimation algorithms typically fail to provide trustworthy  uncertainty quantification \citep{wang2019frequentist}.
Additionally, the parameters of interest for CCA $(\rho, \mathbf{u}^*, \mathbf{v}^*)$ are complex functional of the model parameters $(\mathbf{A}^{(m)}, \mathbf{\Phi}^{(m)})$, and so the impact of the approximation to the posterior of $(\mathbf{A}^{(m)}, \mathbf{\Phi}^{(m)})$ relative to the posterior of the CCA parameters may be unclear.

\pagebreak 

\begin{appendices}
\section{Shrinkage Properties of the Multiplicative Half-Cauchy Process Prior}
In this section, we consider a few theoretical results regarding the behavior of our Multiplicative Half-Cauchy Prior.
Following the framework of \citet{durante2017note}, we focus on the 
properties of the model imposed on the shrinkage parameters $(\eta_1,\ldots,\eta_k,\ldots,\eta_d)$.  For this discussion, we let MHCP denote the Multiplicative Half-Cauchy  Prior, and consider $(\eta_1,\ldots,\eta_d)\sim\mathrm{MHCP}(\zeta)$ as defined previously.  For one realization, we have  $\eta_k=\prod_{j=1}^k \tilde{\eta}_j$, where $\tilde{\eta}_1\sim C^+(0,1)$ and $\tilde{\eta}_k\sim C^+(0,\zeta)$ for $j=2,\ldots,d$.

In the spirit of Proposition 1 by \citet{durante2017note}, we first show that our prior structure has full support on the $d$-dimensional half-line.  We consider any $(\eta^{0}_1,\ldots,\eta^{0}_d)\in (0,\infty)^d$, and we wish to show that MHCP places positive probability on the neighborhood around $(\eta^{0}_1,\ldots,\eta^{0}_d)$.  That is, for any $\epsilon>0$, $P\left\{ \sum_{k=1}^d |\eta_k - \eta^0_k| < \epsilon\right\}>0$.  Following the same steps as \citet{durante2017note}, this result is trivial since each $\tilde{\eta}_k$ has full support on $(0,\infty)$.

More importantly, we are able to show that MHCP places increasing mass in neighborhoods of zero as we allow $k$ to increase.  This key result is what guarantees that MHCP can effectively shrink unnecessary coefficients by decreasing the impact of each subsequent column of the loading matrix.  In the spirit of Lemma 2 from \citet{durante2017note}, we developed a lemma stated and proved below. 

Prior to considering the lemma,  we require a preliminary result.

\noindent {\bf Result.}  $W_1\sim \text{C}^+(0,a)$ and $W_2\sim \text{C}^+(0,b)$ independent.  Then the distribution for $U=W_1 W_2$ is
\begin{equation}
f(u) = \frac{2ab}{\pi^2}\,\frac{\log(u^2/a^2 b^2)}{(u^2-a^2b^2)}, \ u\in \mathbb{R^+}.
\label{eq:HCprod}
\end{equation}

\begin{proof}
    
The joint distribution of $(W_1,W_2)$ is
\begin{equation*}
f(w_1,w_2) = \frac{1}{\pi^2}\,\frac{4ab}{(w_1^2+a^2) (w_2^2+b^2)}, \quad \ w_1, w_2 \in \mathbb{R^+}.
\end{equation*}
The distribution follows from  standard results regarding the distribution of a  product of random variables \citep[e.g.,][]{rohatgi2015introduction} and a partial fraction decomposition.  For $u>0$, the PDF for $U=W_1 W_2$ is
\begin{eqnarray*}
f(u) &=& \int f_{W_1,W_2}(w_1, u/w_1)\frac{1}{w_1}\,dw_1 \\
&=& \frac{4ab}{\pi^2} \int_{0}^{\infty}\frac{w_1}{(w_1^2+a^2)(u^2+w_1^2b^2)} \ dw_1 \\
&=&\frac{4ab}{\pi^2(u^2 -a^2b^2)} \int_{0}^{\infty}  \left[\frac{w_1} {w_1^2+a^2} -\frac{b^2 w_1}{u^2+w_1^2b^2}\right]\ dw_1 \\
&=& \frac{4ab}{\pi^2(u^2 -a^2b^2)} \lim_{W\rightarrow\infty} \left\{ \frac{1}{2} \left[ \log(w_1^2+a^2)-\log(u^2+w_1^2b^2) \right|_{w_1=0}^W \right\} \\
&=& \frac{4ab}{\pi^2(u^2 -a^2b^2)}  \lim_{W\rightarrow\infty} \left\{ \frac{1}{2} \left[ \log(W^2+a^2)-\log(u^2+W^2b^2) - \log(a^2) + \log(u^2) \right] \right\} \\
&=& \frac{4ab}{\pi^2(u^2 -a^2b^2)}  \lim_{W\rightarrow\infty}\left\{ \frac{1}{2}  \log\left(\frac{u^2(W^2+a^2)}{a^2(u^2+W^2b^2)}\right)  \right\} \\
&=& \frac{4ab}{\pi^2(u^2 -a^2b^2)}  \left\{ \frac{1}{2}  \log\left(\frac{u^2}{a^2b^2}\right)  \right\} \\
&=& \frac{2ab}{\pi^2}\,\frac{\log(u^2/a^2 b^2)}{(u^2-a^2b^2)},
\end{eqnarray*}
as required.



\end{proof}
Now, we state the lemma we desire to prove.  \\
\noindent {\bf Lemma 1.} Consider $(\eta_1,\ldots,\eta_d)\sim\mathrm{MHCP}(\zeta)$.  For all $k=1,\ldots,d-1$, $P\left\{\eta_{k+1} \leq t\right\} \geq P\left\{ \eta_k \leq t\right\}$ for all $t$ sufficiently close to zero.

\begin{proof} 
Firstly, we consider the conditional distributions of $\eta_k$ and $\eta_{k+1}$, given $\eta_{k-1}$.  When $k=1$, we define $\eta_0=1/\zeta$ for completeness.  From scale family results, we have $\eta_k\mid \eta_{k-1}\sim C^+(0,\zeta\eta_{k-1})$, and hence
\begin{equation*}
     f_{\eta_{k}|\eta_{k-1}}(t\mid \eta_{k-1})=  \frac{2}{\pi}\,\frac{\zeta\eta_{k-1}}{t^2+\zeta^{2}\eta_{k-1}^{2}}, \quad t\in \mathbb{R^+}.
 \end{equation*}
To get the distribution of $\eta_{k+1}$ given $\eta_{k-1}$, we apply \eqref{eq:HCprod} to  $W_1=\eta_k=\tilde{\eta}_{k}\eta_{k-1}\sim \text{C}^+(0,a)$, $a=\zeta\eta_{k-1}$, and $W_2=\tilde{\eta}_{k+1}\sim \text{C}^+(0,b)$, $b=\zeta$.
The conditional distribution of $\eta_{k+1}=\eta_{k-1}\tilde{\eta}_{k}\tilde{\eta}_{k+1}=W_1 W_2$ is given by 
\begin{equation}
 f_{\eta_{k+1}|\eta_{k-1}}(t\mid \eta_{k-1})= \frac{2\zeta^{2}\eta_{k-1}}{\pi^2}\,\frac{\log(t^2/\zeta^{4}\eta_{k-1}^{2})}{(t^2-\zeta^{4}\eta_{k-1}^{2})}, \quad t\in \mathbb{R^+}. 
 \nonumber
 \end{equation}
Consequently, 
\begin{equation}\label{eqn:ratio}
    \frac{f_{\eta_{k+1}|\eta_{k-1}}(t\mid \eta_{k-1})}{f_{\eta_{k}|\eta_{k-1}}(t\mid \eta_{k-1})}= \frac{\eta_{k-1}}{\pi} \frac{\log(t^2/\zeta^{4}\eta_{k-1}^{2} ) (\zeta^2\eta_{k-1}^2+t^2)}{(t^2-\zeta^{4}\eta_{k-1}^{2})}, \ t\in \mathbb{R}^{+}.
\end{equation} 
Now, we will study limiting properties of \eqref{eqn:ratio}, as $t \rightarrow 0^{+}$. Note that 
\begin{eqnarray}
\lim_{t \to 0^{+}} \frac{f_{\eta_{k+1}|\eta_{k-1}}(t\mid \eta_{k-1})}{f_{\eta_{k}|\eta_{k-1}}(t\mid \eta_{k-1})} &=& \nonumber   \frac{\eta_{k-1}}{\pi} \lim_{t \to 0^{+}} \frac{\log(t^2/\zeta^{4}\eta_{k-1}^{2} ) (\zeta^2\eta_{k-1}^2+t^2)}{(t^2-\zeta^{4}\eta_{k-1}^{2})}, \\ 
&=& \frac{\eta_{k-1}}{\pi} 
\left\{  \lim_{t \to 0^{+}} \left[ \log(t^2/\zeta^{4}\eta_{k-1}^{2} )\right] \times  \lim_{t \to 0^{+}} \left[\frac{\zeta^2\eta_{k-1}^2+t^2}{t^2-\zeta^{4}\eta_{k-1}^{2}} \right] \right\}.\nonumber \\
&=& \frac{\eta_{k-1}}{\pi} 
\left\{  \left[ -\infty \right] \times   \left[\frac{1}{-\zeta^{2}} \right] \right\} = \infty.\nonumber
\end{eqnarray}
That is,  the conditional distribution of $\eta_{k+1}$ places increasingly greater mass at values near zero than the conditional distribution of $\eta_k$ (with both distributions conditional on the same $\eta_{k-1}$).  Obviously, this implies $f_{\eta_{k+1}|\eta_{k-1}}(t\mid \eta_{k-1})\geq f_{\eta_{k}|\eta_{k-1}}(t\mid \eta_{k-1})$ for all $t$ sufficiently small.
The remainder of the proof follows the same argument as in Lemma 2 of \citet{durante2017note}.  In particular, for $t$ sufficiently small, the ordering of the conditional densities for $\eta_{k+1}$ and $\eta_k$ will be maintained in the marginal PDFs: 
\begin{eqnarray*}
f_{\eta_{k+1}}(t ) &=& \mathrm{E}_{\eta_{k-1}}\left[ f_{\eta_{k+1}|\eta_{k-1}}(t\mid \eta_{k-1})\right] \geq \mathrm{E}_{\eta_{k-1}}\left[ f_{\eta_{k}|\eta_{k-1}}(t\mid \eta_{k-1})\right] = f_{\eta_{k}}(t ) .
\end{eqnarray*}
This clearly implies $P\left\{\eta_{k+1} \leq t\right\} \geq P\left\{ \eta_k \leq t\right\}$ as in the statement of Lemma 1.
\end{proof}

This result shows that the MHCP does in fact favor sparsity in the columns of the loading matrices $\mathbf{A}^{(m)}$ a priori.  We also consider the flexibility and adaptability of our prior in terms of the following additional lemma.\\
\noindent {\bf Lemma 2.} Consider $(\eta_1,\ldots,\eta_d)\sim\mathrm{MHCP}(\zeta)$.  For all $k=1,\ldots,d-1$, $P\left\{\eta_{k+1} \geq t\right\} \geq P\left\{ \eta_k \geq t\right\}$ for all $t$ sufficiently large.

\begin{proof} 
The proof is basically identical to the proof of Lemma 1 by showing 
\[ \lim_{t \to \infty } \frac{f_{\eta_{k+1}|\eta_{k-1}}(t\mid \eta_{k-1})}{f_{\eta_{k}|\eta_{k-1}}(t\mid \eta_{k-1})} =\infty,\]
which provides $f_{\eta_{k+1}}(t ) \geq  f_{\eta_{k}}(t )$ for all large $t$.
\end{proof}
This implies that our prior structure is protected from over-shrinkage as it places continually larger mass in the tails of the distribution of the column shrinkage parameter.  This is, of course, a consequence of using the heavy-tailed Half-Cauchy distribution for the multiplicative adjustment $\tilde{\eta}$.  In contrast to the more common multiplicative gamma process \citep{bhattacharya2011sparse, durante2017note}, we expect our MHCP to be able to adapt more easily to differences in the magnitudes of the columns in $\mathbf{A}^{(m)}$ as it considers higher density on the edges of the distribution.

In conjunction, Lemmas 1 and 2 tells us that the MHCP is placing increasing mass at very small and very large values of $\eta_k$.  However, in order for this to be an effective shrinkage structure, it is critical that it is shifting  mass towards zero faster than it is shifting mass to  favor large values.  To see that this is the case, we rely on one final lemma.\\
{\bf Lemma 3.} Consider $(\eta_1,\ldots,\eta_d)\sim\mathrm{MHCP}(\zeta)$.  For $k=1,\ldots,d$, the prior median of $\eta_k$ is $\zeta^{k-1}$.
\begin{proof}
For $k=1$, the distribution of $\eta_1$ is $\text{C}^+(0,1)$, which clearly has median $\zeta^0=1$.  

For $W_k\sim\text{C}^+(0,a_k)$, the distribution of $U_k=\log(W_k/a_k)$ is 

\begin{equation}\label{eq:hsd}
    f(u_k) = \frac{2}{\pi},\frac{1}{e^{-u}+e^{u}}, \quad u \in\mathbb{R}, 
\end{equation}

which is clearly a symmetric distribution, and will have a median of 0.  Note that the distribution \eqref{eq:hsd} is closely related to the hyperbolic secant distribution, although it is missing a scaling term $\pi$ \citep{harkness1968generalized,ding2014three}.
It follows that the distribution of $\log(W_k)= U_k +\log(a_k)$ will  be symmetric around its median $\log(a_k)$.  As the distribution of the sum of independent symmetric variables will also be symmetric, it follows that $\sum_{j=1}^k U_j$ is symmetric with mean zero.  Consequently, $\sum_{j=1}^k \log(W_j)$ will have median $\sum_{j=1}^k \log(a_j)$.

To obtain the median for $k=2,\ldots, d$, we using the above notation by  characterizing MHCP through $W_k=\tilde{\eta}_k\sim\text{C}^+(0,a_k)$ for all $k=1,\ldots,d$, with $a_1=1$ and $a_k=\zeta$ ($k\neq 1$).  Then, considering $\eta_k=\prod_{j=1}^k \tilde{\eta}_j$ for $k>1$, it follows that $\log(\eta_k) = \sum_{j=1}^k \log(\tilde{\eta}_j) = \sum_{j=1}^k \log(W_j)$ will have median $\sum_{j=1}^k \log(a_k) = (k-1)\log(\zeta)$.  As the median is preserved under monotonic transformations, it follows that the median of $\eta_k$ will be $\zeta^{k-1}$.
\end{proof}
Hence, for all $\zeta<1$, the median of $\eta_k$ will be decreasing towards zero, providing increasing shrinkage of the columns of $\mathbf{A}^{(m)}$.


\section{Further Simulation Results}
In these Supplementary Material, we include further simulation results that we have not included in Section 5 of the main manuscript.

\subsection{Simulation Results: Diagonal and Non-diagonal Model Selection}\label{Subsection:Overshrinkage_effects}
\begin{table}[tbp]
  \centering
  \caption{Proportion (\%) of Data Sets With Potential Overshrinkage}
    \begin{tabular}{cccccccc}
    \hline
    Setting &  1     & 2     & 3     & 4     & 5     & 6  & 7 \\
    \cline{2-8}
   Proportion with $P(\rho<0.2)> 0.5$ & 0     & 14    & 0     & 0     & 0     & 1  & 0 \\
    \hline
     \end{tabular}%
  \label{Table:Over_Shrinkage}%
\end{table}%
In this section we investigate our combined model selection strategy.  Recall that our combined strategy switches from the NDFSM model results to the DFSM if there is evidence of overshrinkage based on the posterior probability that $\rho_1<0.2$. If this posterior probability exceeds 0.5, then we use the diagonal model results.  
Table \ref{Table:Over_Shrinkage} shows proportion of data sets when this criteria is met, that is, when the NDFSM+DFSM strategy bases inference on the more restricted DFSM. 
\par

As noted previously, this only appears when we consider cases with more features than observations, representing cases 2 and 6.  When it does occur, it is fairly rare, impacting only 14\% of cases in setting 2 and 1 of the 100 cases in setting 6. When the true generalized specificity matrices are diagonal (case 4), we also do not observe any overshrinking cases.  
\par 

\begin{table}[tb]
  \centering
  \caption{RMSE Comparison by Overshrinking Criteria}
    \begin{tabular}{ccc}
    \hline
    Method & \multicolumn{2}{c}{Setting 2} \\
    & Potentially Overshrunk & Not Overshrunk \\ \cline{2-3}
  $N$ & $N=14$ & $N=86$ \\ 
    NDFSM & 0.5674 & 0.1011 \\
    DFSM  & 0.1626 & 0.0620 \\
    GFA   & 0.4644 & 0.2389 \\
    GHS   & 0.0756 & 0.0537 \\
  \hline
 & \multicolumn{2}{c}{Setting 6} \\
    & Potentially Overshrunk & Not Overshrunk \\  \cline{2-3}
  $N$ & $N=1$ & $N=99$ \\ 
    NDFSM & 0.5026 & 0.1085 \\
    DFSM  & 0.1243 & 0.0434 \\
    GFA   & 0.0714 & 0.2293 \\
    GHS   & 0.0842 & 0.0530 \\
  \hline
    \end{tabular}%
  \label{Table:RMSE_Change_Overshrinkage}%
\end{table}%


In Table \ref{Table:RMSE_Change_Overshrinkage} we further investigate estimation error for 1st CC, stratifying by these suspected overshrunk outputs compared to the remaining estimates unaffected by overshrinking. 
Viewing the NDFSM rows, there is a clear difference in the RMSE between those cases whether the estimates are flagged as overshrunk vs not.  To help with comparison, we include the DFSM, GFA and GHS results, also stratified by the NDFSM overshrinking criteria.  For the unshrunk cases, NDFSM estimates the first CC with similar accuracy to DFSM and GHS, although it is slightly worse (GFA is consistently poor in these cases).  The missestimation with NDFSM in Setting 2 is clearly dominated by the poor performance in these 14 datasets with overshrinking, and our combined strategy will replace these poor estimates with the DFSM estimates that are much more accurate.  Similar behavior is occuring in Setting 6, but it is less dramatic since there is only a single data replication with overshrinkage.
\par

 \subsection{Simulation Results: Significant Variable Loadings}\label{Section:Significant_Variable_Loading_supp}
In conjunction with the previous exploration of the accuracy of the canonical loading vectors $\hat{\mathbf{u}}^*$ and $\hat{\mathbf{v}}^*$, we also want to interrogate whether we are able to correctly detect whether a variable is significantly loaded or not.  
Recall that for Bayesian variable selection we consider a variable to be significantly associated with the CC direction if the 50\% credible interval for its factor loading excludes zero. In the penalized methods, SCCA (O) and SCCA (STD), if the component in the estimated (sparse)  CC direction is non-zero, then we say that the corresponding variable is significantly contributing to the CC calculation. 

To characterize the true effect of each variable, we divide the elements of true data-generating direction vectors into 3 groups according to their contribution to the calculation of CC and direction vectors. Features are considered relevant in both the  latent and CCA structure, if they have non-zero value in the column of the factor loading matrix $\mathbf{A}^{(m)}$ producing the first canonical correlation.  In our data generating approach,  there are 5 such features. 
Recall that variables unrelated to the latent factor structure can be loaded on the canonical correlation if they are highly associated through $\mathbf{\Phi}^{(m)}$ with another variable with non-zero projection value.
As the direction vectors are the complex function of $\mathbf{\Sigma}$, it is difficult to understand the direct impact of the AR structure in the generalized specificity on the estimands. To that end, we define further features that are relevant in the  CCA structure if their true loading value is greater than 0.1 in absolute value even if the latent factor $\mathbf{z}$ is not associated with the views; we consider the threshold 0.1 as  that indicates the squared contribution is greater than 0.01 (or 1\%) of the total direction vector. 
There are 9 such features in the settings with an AR structure for $\mathbf{\Phi}^{(m)}$, and no such features in settings 3 and 4 with diagonal $\mathbf{\Phi}^{(m)}$. If the absolute factor loading is less than 0.1, then it is practically irrelevant, so it goes in the third block of features unrelated to the CC.  There are 136 elements in the AR settings and 145 in the independence settings. The better performing methods are those  which have high selection rates in block 1 (and to a lesser extent, block 2) and low selection rates in block 3. Due to the low effect sizes of the factor loadings in block 2, we expect lower variable selection rates than in block 1. \par

\begin{table}[tb]
  \centering
  \caption{Percentage Accuracy of Significant Variable Loading }
     \begin{tabular}{cccccccc}
     \hline
      & \multicolumn{7}{c}{Settings } \\
      Methods & 1     & 2     & 3     & 4     & 5     & 6     & 7 \\
    \hline
          & \multicolumn{7}{c}{Features Relevant to Latent and CCA Structure} \\
    \cline{2-8}
    NDFSM & 100.00 & 86.4 & 100 & 99.6 & 100.00 & 83.6 & 99.2 \\
    DFSM  & 100.00 & 97.80 & 100.00 & 99.80 & 100.00 & 94.00 & 89.20 \\
    GHS   & 41.80 & 39.80 & 40.40 & 39.20 & 42.60 & 37.00 & 46.20 \\
    GFA   & 100.00 & 99.80 & 100.00 & 100.00 & 100.00 & 73.00 & 98.00 \\
    SCCA (STD) & 82.60 & 36.80 & 81.00 & 32.60 & 76.20 & 29.80 & 68.60 \\
    SCCA (O) & 46.60 & 24.60 & 42.20 & 31.40 & 45.60 & 25.60 & 63.80 \\
    \hline
          & \multicolumn{7}{c}{Features Relevant to CCA Structure} \\
     \cline{2-8}
    NDFSM & 82.13 & 10.63 & -    & -    & 82.50 & 8.38 & 81.00 \\
    DFSM  & 23.62 & 3.63  & -    & -    & 21.75 & 2.63  & 26.00 \\
    GHS   & 14.75 & 2.15  & -    & -    & 15.75 & 3.62  & 21.25 \\
    GFA   & 48.75 & 55.25 & -    & -    & 41.50 & 65.50 & 83.25 \\
    SCCA (STD) & 0.00  & 0.25  & -    & -    & 0.00  & 0.00  & 0.00 \\
    SCCA (O) & 0.00  & 0.62  & -    & -    & 0.00  & 0.75  & 0.38 \\
     \hline
          & \multicolumn{7}{c}{Non Relevant Features} \\
     \cline{2-8}
    NDFSM & 5.28  & 0.96  & 0.00  & 1.78 & 5.12  & 1.37  & 4.39 \\
    DFSM  & 3.52  & 2.07  & 1.20  & 2.76  & 4.05  & 2.36  & 3.04 \\
    GHS   & 3.65  & 0.66  & 0.02  & 3.09  & 4.92  & 2.88  & 2.38 \\
    GFA   & 43.69 & 53.79 & 46.73 & 47.55 & 37.42 & 63.70 & 80.84 \\
    SCCA (STD) & 0.00  & 0.23  & 0.00  & 0.26  &0.03  & 0.36  & 0.00 \\
    SCCA (O) & 0.00  & 0.36  & 0.00  & 0.56  & 0.03  & 0.18  & 0.04 \\
     \hline
    \end{tabular}%
  \label{Table:Variable selection}%
\end{table}%

Table \ref{Table:Variable selection} summarizes the results. We can see that in block 1 our methods all consistently recover the true features involved in the latent structure.  While GFA also performs well in block 1, GHS selects these most critical variables less than 50\% of the time across all methods.  This is 
consistent with the poorer results on direction vector recovery. The penalized approaches perform decently in the low-dimensional settings but not in high-dimensional cases.   
\par 
In block 2, NDFSM continues to have almost perfect recovery in the lower-dimension ($n=300$) settings, with a substantial drop off when $n=50$.  As would be anticipated, DFSM does worse in this block as these are the features whose role is governed by the non-diagonal specificity matrix which DFSM misspecifies. 
GFA tends to do fairly well in this block typically identifying 40--60\% of features, which is worse than NDFSM for small dimension but better for large.
GHS has a low selection rate as in block 1. Here penalized methods are unable to identify features which result from the AR correlations. \par

In block 3 we expect that selection rates should be low for all models, but GFA shows a high false positive rate in this case.  In fact, it is selecting completely irrelevant variable at approximately the same rate it was selectioning feature in block 2.  Our FSM models seem to control the rate of false discoveries, and consistent with their low power throughout, GHS has low selection rates. Penalized methods performed the well in this block as they rarely not pick any features which are not relevant to the CC structure. 
\par

\end{appendices}

\bibliographystyle{elsarticle-harv} 
\bibliography{Citation}

\end{document}